\documentclass[%
reprint,
superscriptaddress,
showpacs,
nofootinbib,
amsmath,amssymb,
prc
notitlepage,
floatfix
]{revtex4-1}

\usepackage{longtable}
\usepackage{graphicx}
\usepackage{dcolumn}
\usepackage{bm}
\usepackage{color}
\usepackage{hyperref}
\definecolor{LinkColor}{rgb}{0,0,0.5}
\hypersetup{colorlinks=true,%
linkcolor=LinkColor,%
citecolor=LinkColor,%
filecolor=LinkColor,%
menucolor=LinkColor,%
pagecolor=LinkColor,%
urlcolor=LinkColor}
\usepackage[caption=false, labelformat=empty]{subfig}
\usepackage{microtype}
\usepackage{comment}

\long\def\symbolfootnote[#1]#2{\begingroup\def\thefootnote{\fnsymbol{footnote}}\footnote[#1]{#2}\endgroup} 
\begin{document}


\title{Measurement of the reaction $^{17}$O$(\alpha,n)^{20}$Ne and its impact on the $s$ process in massive stars}

\author{A. Best}
\altaffiliation{Present address: Lawrence Berkeley National Laboratory, Berkeley, CA 94720, USA}
\email{abest1@nd.edu}
\affiliation{Department of Physics, University of Notre Dame, Notre Dame, IN 46556, USA}
\author{M. Beard}
\altaffiliation{NuGrid Collaboration, \url{http://www.nugridstars.org}}
\affiliation{Department of Physics, University of Notre Dame, Notre Dame, IN 46556, USA}
\affiliation{ExtreMe Matter Institute EMMI, GSI Helmholzzentrum f\"{u}r Schwerionenforschung, Plankstra\ss e 1, 54291, Darmstadt, Germany}
\author{J. G\"{o}rres}
\affiliation{Department of Physics, University of Notre Dame, Notre Dame, IN 46556, USA}
\author{M. Couder}
\author{R. deBoer}
\affiliation{Department of Physics, University of Notre Dame, Notre Dame, IN 46556, USA}
\author{S. Falahat}
\affiliation{Department of Physics, University of Notre Dame, Notre Dame, IN 46556, USA}
\affiliation{Department for Biogeochemistry, Max-Planck-Institute for Chemistry, 55020 Mainz, Germany}
\author{R. T. G\"uray}
\affiliation{Kocaeli University, Department of Physics, Umuttepe 41380, Kocaeli, Turkey}
\author{A. Kontos}
\altaffiliation{Present address: National Superconducting Cyclotron Laboratory, Michigan State University, East Lansing, MI 48824, USA}
\affiliation{Department of Physics, University of Notre Dame, Notre Dame, Indiana 46556, USA}
\author{K.-L. Kratz}
\affiliation{Department for Biogeochemistry, Max-Planck-Institute for Chemistry, 55020 Mainz, Germany}
\author{P. J. LeBlanc}
\altaffiliation{Present address: CANBERRA Industries Inc., Meriden, CT 06450}
\author{Q. Li}
\author{S. O'Brien}
\altaffiliation{Present address: US Government, Washington DC 20009}
\affiliation{Department of Physics, University of Notre Dame, Notre Dame, IN 46556, USA}
\author{N. \"Ozkan}
\affiliation{Kocaeli University, Department of Physics, Umuttepe 41380, Kocaeli, Turkey}
\author{M. Pignatari}
\altaffiliation{NuGrid Collaboration, \url{http://www.nugridstars.org}}
\affiliation{Department of Physics, University of Basel, Basel, 4056, Switzerland}
\author{K. Sonnabend}
\affiliation{Institute for Applied Physics, Goe\-the-Uni\-ver\-si\-ty Frankfurt, 60325 Frankfurt, Germany}
\author{R. Talwar}
\author{W. Tan}
\author{E. Uberseder}
\author{M. Wiescher}
\affiliation{Department of Physics, University of Notre Dame, Notre Dame, IN 46556, USA}

\date{\today}

\begin{abstract}
\begin{description}
\item[Background] The ratio between the rates of the reactions $^{17}$O$(\alpha,n)^{20}$Ne and $^{17}$O$(\alpha,\gamma)^{21}$Ne determines whether $^{16}$O is an efficient neutron poison for the $s$ process in massive stars, or if most of the neutrons captured by $^{16}$O$(n,\gamma)$ are recycled into the stellar environment. This ratio is of particular relevance to constrain the $s$ process yields of fast rotating massive stars at low metallicity.

\item[Purpose] Recent results on the $(\alpha,\gamma)$ channel have made it necessary to measure the $(\alpha,n)$ reaction more precisely and investigate the effect of the new data on $s$ process nucleosynthesis in massive stars.

\item[Method] The $^{17}$O$(\alpha, n_{(0+1)})$ reaction has been measured with a moderating neutron detector. In addition, the $(\alpha, n_{1})$ channel has been measured independently by observation of the characteristic  1633~keV $\gamma$-transition in $^{20}$Ne. The reaction cross section was determined with a simultaneous \emph{R}-matrix fit to both channels. $(\alpha,n)$ and $(\alpha, \gamma)$ resonance strengths of states lying below the covered energy range were estimated using their known properties from the literature.

\item[Results] 
The reaction channels $^{17}$O$(\alpha,n_0)^{20}$Ne and $^{17}$O$(\alpha,n_1\gamma)^{20}$Ne were measured in the energy range E$_{\alpha} = 800$~keV to 2300~keV. A new $^{17}$O$(\alpha,n)$ reaction rate was deduced for the temperature range 0.1~GK to 10~GK. At typical He burning temperatures, the combination of the new $(\alpha,n)$ rate with a previously measured $(\alpha,\gamma)$ rate gives approximately the same ratio as current compilations. The influence on the nucleosynthesis of the $s$ process in massive stars at low metallicity is discussed.

\item[Conclusions] It was found that in He burning conditions the $(\alpha,\gamma)$ channel is strong enough to compete with the neutron channel. This leads to a less efficient neutron recycling compared to a previous suggestion of a very weak $(\alpha,\gamma)$ channel. $S$ process calculations using our rates confirm that massive rotating stars do play a significant role in the production of elements up to Sr, but they strongly reduce the $s$ process contribution to heavier elements.
	
\end{description}
\end{abstract}

\maketitle
\section{INTRODUCTION}
Most of the elements in the mass range 60 $<$ A $<$ 90 that we observe today in the Solar system are produced by neutron capture on iron seed nuclei, mainly during the convective
core helium and convective shell carbon phases in massive stars (weak $s$ process) \cite{Kaeppeler:2011}. The efficiency of the weak $s$ process depends on the network of nuclear reactions used
for stellar calculations, above all on the main neutron source $^{22}$Ne$(\alpha,n)^{25}$Mg and on its main competing reaction $^{22}$Ne$(\alpha,\gamma)^{26}$Mg, and
on the set of neutron capture cross sections used in the energy range of He and C burning (0.25 $\lesssim$ T $\lesssim$ 1.5 GK) \cite{Nassar:2005, Heil:2008b, Pignatari:2010}.

Since the rate of neutron captures is slow compared to the decay rate of unstable reaction products, in the $s$ process the neutron capture path follows the valley of stability. The final $s$ process abundances depend on the total amount of neutrons available integrated over time (or neutron exposure)~\cite{Clayton:1968}, the history of the neutron density in the stellar regions where the $s$ process occurs, and on the initial stellar metallicity. Light isotopes, depending on their abundance and neutron capture cross sections, can capture a large amount of free neutrons, thereby acting as neutron poisons in the burning environment.

The main neutron source for the $s$ process in massive stars is the $^{22}$Ne$(\alpha,n)^{25}$Mg reaction. It is activated mostly at the end of the convective core He burning \cite{Kaeppeler:1994} and
in the convective carbon shell, on ashes of previous He core burning \cite{Raiteri:1991}. The abundance of $^{22}$Ne is given by the initial CNO abundances, where $^{14}$N produced via
H burning in the CNO cycle forms $^{22}$Ne by capturing two $\alpha$-particles in the initial He burning phases (\cite{The:2007, Pignatari:2010} and references therein).
The $s$ process in massive stars is a secondary process because of its dependence on the initial metallicity, the $^{22}$Ne abundance, on the iron seeds and on the effect
of light neutron poisons. 

Because the efficiency and yields of the $s$ process decrease linearly with the initial metal content for massive stars with lower metallicities \cite{Woosley:1995, Raiteri:1992, Baraffe:1992}, the $s$ process contribution to galactic chemical evolution from massive stars becomes marginal at low metallicity. However, recent observations (e.g.,~\cite{Truran:2002, Aoki:2005, Aoki:2006, Chiappini:2011}) show a puzzling enhancement of light $s$ process elements in very low metallicity stars. Recent theoretical studies have shown that fast rotating massive stars may potentially have $s$ process yields orders of magnitude higher than non-rotating stars~\cite{Pignatari:2008, Frischknecht:2012}. In this case primary $^{22}$Ne is produced in the convective He core, independently from the initial metallicity of the star. As a result, fast rotator yields could be relevant for the chemical evolution of heavy $s$ process elements. 

Baraffe et al. \cite{Baraffe:1992} discussed the potential relevance of the $^{17}$O$(\alpha,\gamma)$ and $^{17}$O$(\alpha,n)$ rates for the $s$ process in massive stars at low metallicity. Specifically, the relative $(\alpha,n)/(\alpha,\gamma)$ ratio constrains the neutron poison efficiency of the $^{16}$O$(n,\gamma)^{17}$O reaction. $^{16}$O is the most abundant species in regions processed by advanced He burning via the $^{12}$C$(\alpha,\gamma)^{16}$O reaction. Because of its high abundance $^{16}$O is the strongest neutron absorber, despite the low $^{16}$O$(n,\gamma)^{17}$O cross section \cite{Pignatari:2010}. The present large uncertainties of the $^{17}$O+$\alpha$ rates (together with other relevant rates, e.~g., $^{22}$Ne+$\alpha$~\cite{Pignatari:2008}) affects the theoretical $s$ process predictions of fast rotating massive stars at low metallicity~\cite{Hirschi:2008}. 

The neutron source $^{22}$Ne$(\alpha,n)^{25}$Mg is activated in massive stars at stellar temperatures of typically T $\sim$ 0.25~--~0.3 and T $\sim$ 1~GK, during convective core He burning and convective shell C burning, respectively. For the reaction $^{17}$O$+\alpha$, center-of-mass energies of around 0.5~MeV and 1.1~MeV are therefore most important. The reaction $^{17}$O$(\alpha,n)^{20}$Ne has previously been
measured in the energy range between 0.6 and 12.5~MeV \cite{Denker:1994,Bair:1973,Hansen:1967}. The lower energy neutron data are only available in form of a Ph.D. thesis \cite{Denker:1994}. Its reliability is uncertain due to a strong background
contribution from the $^{18}$O$(\alpha,n)^{21}$Ne channel, as well as the inability of the experimental setup to discriminate between the n$_0$ and n$_1$ channels, leading to
ambiguities in the detector efficiency determination.

To overcome these problems we remeasured the total yield of the reaction $^{17}$O$(\alpha,n_{\text{total}})^{20}$Ne observed using a high efficiency neutron detector and targets with a very
low (0.4\%) ${}^{18}$O content. The experimental setup and the measurement are described in sections \ref{sec:n_total_setup} and \ref{sec:ntot_results}, respectively. In an
additional experiment the $^{17}$O$(\alpha,n_1 \gamma)^{20}$Ne channel was measured by observation of the E = 1633~keV $\gamma$-transition to the ground state of ${}^{20}$Ne
(sections \ref{sec:n_1_setup} and \ref{sec:n1_results}). This allowed a subtraction of the n$_1$ contribution to the total yield and a better determination of the effective neutron
detector efficiency. In section \ref{sec:r-matrix} we present a simultaneous \emph{R}-matrix fit to both neutron data sets.

Until recently the only available rates for the competing $^{17}$O$(\alpha,\gamma)^{21}$Ne channel were an estimate from the Caughlan-Fowler compilation \cite{CF88} (CF88 hereafter) 
and a rate on the basis of microscopic cluster model calculations \cite{Descouvemont:1993}. These two rates differ by a factor of $\approx 1000$, causing a large uncertainty in the $s$ process yields from fast rotating massive stars~\cite{Hirschi:2008}. A first measurement of the $^{17}$O$(\alpha,\gamma)^{21}$Ne reaction \cite{Best:2011a} supports the CF88 estimate.
This recent experimental data on the $(\alpha,\gamma)$ channel and the $(\alpha,n)$ data presented here were combined with low-energy extrapolations and used as input for a stellar network calculation. The contributions
to both the $(\alpha,\gamma)$ and $(\alpha,n)$ channels from states below the energy range covered in this experiment
are estimated in section \ref{sec:extrap}. The new reaction rates and a discussion of the astrophysical implications can be found in section \ref{sec:astro}.

\section{EXPERIMENTAL SETUP}
The experiment consisted of two independent measurements, a direct measurement of the reaction neutrons and the detection of the characteristic 1633~keV
$\gamma$-ray from the $^{17}$O$(\alpha,n_1\gamma)^{20}$Ne channel. The specific setups for each part are described separately in the following subsections. Here we describe the parts
of the setup that were common to both measurements.

The $\alpha$-beam was provided by the 4~MV KN accelerator at the University of Notre Dame Nuclear Science Laboratory. 
Energy calibration and resolution (1.1~keV) were determined using the well-known resonances E$_p = 991.86 \pm 0.03$~keV and E$_p = 1317.14 \pm 0.07$~keV in ${}^{27}$Al$(p,\gamma)^{28}$Si \cite{Endt:1990}.
The beam energy was reproducible within $\pm 2$~keV between different hysteresis cycles of the analyzing magnet during the course of the experiment.

Targets were prepared by anodization of 0.3 mm thick tantalum backings using H$_2$O enriched to 90.1\% in $^{17}$O\symbolfootnote[6]{Purchased from Isotech, Miamisburg, OH. The $^{18}$O and $^{16}$O contents of the water were 0.4\% and 9.5 \%, respectively.}.
This process is known to produce homogeneous films of Ta$_2$O$_5$ \cite{Vermilyea:1953,Seah:1988}. The film thickness can be controlled in a reproducible way through regulation
of the maximal anodization voltage. The target thickness was chosen to be about 9~keV of energy loss for an $\alpha$ beam of 1000~keV and a target orientation of $90^{\circ}$ with respect to the beam direction.

To reduce carbon deposition a liquid nitrogen cooled copper tube (cold finger) was mounted in front of the target. A bias of -400 V was applied to the cold
finger for suppression of secondary electrons. The beam was scanned with magnetic steerers to produce a beam spot size of 1.4 cm $\times$ 1.6 cm on the target.
In both parts of the experiment the target chamber was electrically isolated for charge collection and the targets were directly water cooled using deionized water.

\subsection{$^{17}$O$(\alpha, n_1 \gamma)^{20}$Ne setup}\label{sec:n_1_setup}
\begin{figure}[tb]
	\centering
		\includegraphics[width=\columnwidth]{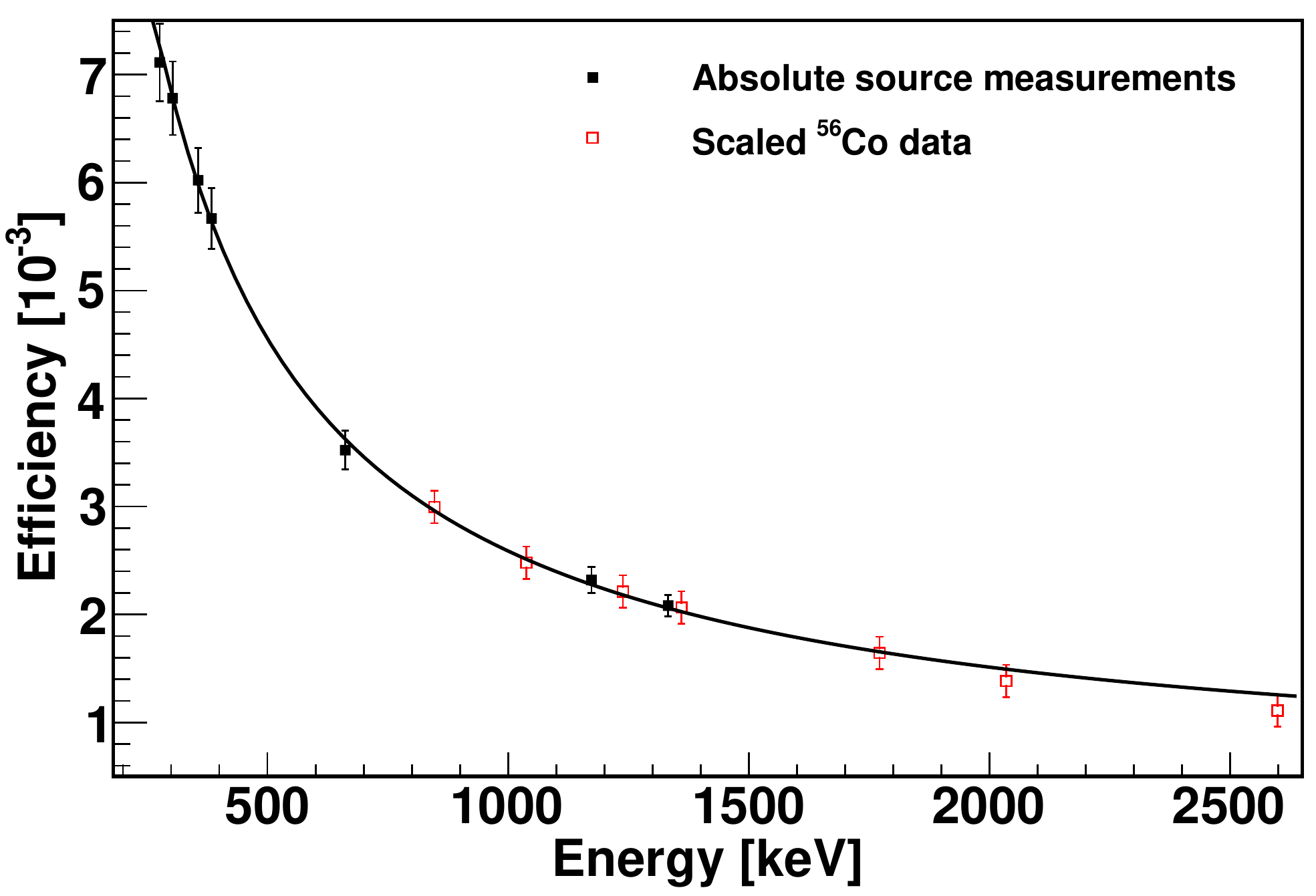}
	\caption{(Color online) Efficiency of the germanium detector as a function of photon energy. The solid line represents a fit to the data points.}
	\label{fig:n1_eff}
\end{figure}

For this part of the measurement a Ge detector with a relative efficiency of 20\% was placed at a distance of 5.4 cm from the target
at an angle of $45^{\circ}$ with respect to the beam direction.
In order to minimize radiation damage to the detector a 2.5 cm polyethylene disk was attached to the detector front cap. This reduced significantly the neutron flux in the detector.
The transition from the first excited state to the ground state of ${}^{20}$Ne emits a $\gamma$ ray with an energy of E$_\gamma = 1633.7$~keV \cite{Tilley:1998}. Absolute efficiencies 
were established with calibrated $^{137}$Cs, $^{60}$Co and $^{133}$Ba sources and augmented using relative efficiency data from $^{56}$Co.
There was enough absorbing material between source and detector (0.25 mm tantalum backing, 1 mm brass target holder and the 2.5 cm thick polyethylene disk)
to attenuate the low-energy photons emitted by the Ba source (E$_{\gamma} = 81$~keV and E$_{\gamma} = 80$~keV) by approximately 98\%, resulting in
negligible summing effects in the Ba measurements. Summing corrections for the ${}^{60}$Co measurements amounted to less than 2\%.
The resulting efficiency curve is shown in Fig. \ref{fig:n1_eff}.

\subsection{$^{17}$O$(\alpha,n_{total})^{20}$Ne setup}\label{sec:n_total_setup}
This part of the experiment was dedicated to the measurement of the $^{17}$O$(\alpha,n_{\text{total}})^{20}$Ne reaction by direct detection of the reaction neutrons. The 
detector consisted of 20 ${}^{3}$He counters that were embedded in an arrangement of two concentric rings with diameter 6~cm and 11~cm, respectively into a 30 cm $\times$ 30 cm $\times$ 33 cm 
polyethylene moderator. The very high cross section of the reaction ${}^{3}$He(n, p)${}^{3}$H (Q = 764~keV, $\sigma = 5330$ barn) for thermal neutron capture and its
low sensitivity to photons make it an excellent neutron detector, although due to the thermalization in the polyethylene moderator the information on the neutron energy is mostly lost.

Additional shielding from background neutrons was provided by a 5 cm thick layer of borated polyethylene on the outside of the moderator. At the beginning of each
experimental run the detector was centered around the target chamber by monitoring the neutron yield as a function of distance of the detector to the target
holder. The position with the maximum yield was then used as the default position for all further measurements. The target was
mounted at an angle of $90^{\circ}$ with respect to the beam direction.
\begin{figure}[tb]
	\centering
		\includegraphics[width=0.7\columnwidth]{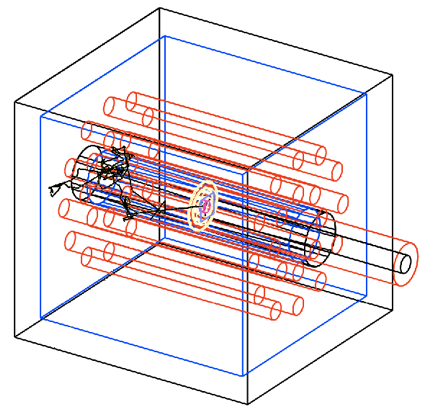}
	\caption{(Color online) Geometry of the neutron detector used in the {\sc Geant4} simulation. Shown are the outlines of the outer borated polyethylene shield, the moderator with the 20 ${}^{3}$He counters and the beam line with the target holder in its center. Depicted in black is the simulated track of a neutron.}
	\label{fig:ndetector_geant}
\end{figure}

\begin{figure}[tb]
	\centering
		\includegraphics[width=\columnwidth]{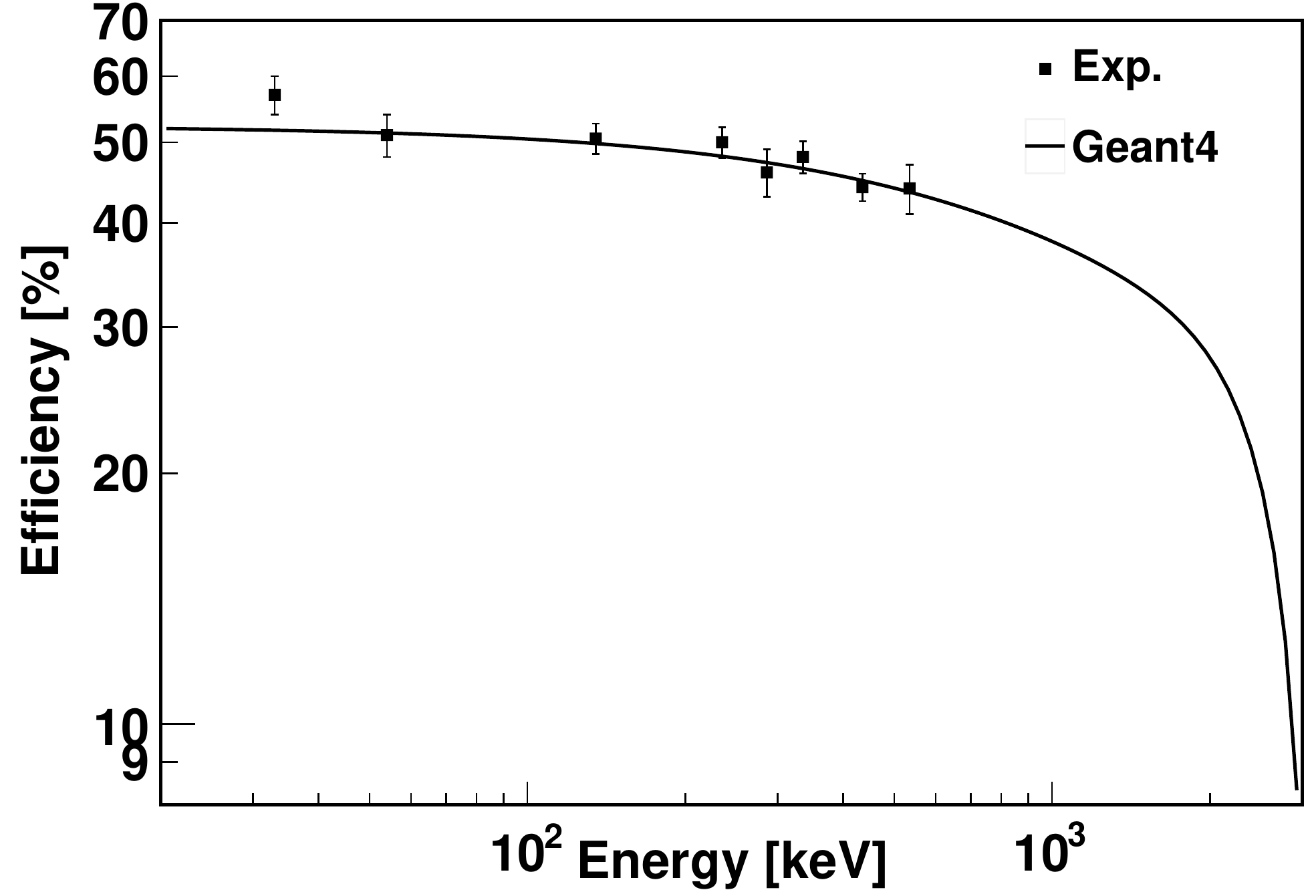}
	\caption{(Color online) Efficiency of the neutron detector as a function of energy. Shown are the results of activation measurements and the {\sc Geant4} simulation. Above a neutron energy of E$_n = 650$~keV the cross section of the ${}^{51}$V(p,n)${}^{51}$Cr reaction is high enough that dead time effects prohibited further measurements. The MCNP data are not shown here but follow the shape of the {\sc Geant4} results.}
	\label{fig:ndetector-eff}
\end{figure}
The absolute detector efficiency was determined using the reaction ${}^{51}$V(p,n)${}^{51}$Cr \cite{Harris:1965, Zyskind:1980, Deconninck:1969}. This reaction with a neutron threshold of 1565~keV is known to produce mono-energetic neutrons that are approximately
isotropic in the center-of-mass system. The thickness of the vanadium targets was about 30 $\mu g / cm^2$, corresponding to a proton energy loss between 2.5 and 3 keV in the covered energy range. Although a large number of narrow resonances has been observed in this reaction
no pronounced angular distributions have been reported in the literature \cite{Gibbons:1955, Salas-Bacci:2004}. Since our data show a smooth behaviour, in agreement with similar calibration measurements using $^{51}$V$(p,n)^{51}$Cr \cite{Ramstrom:1976, Pereira:2010}, we
believe it to be unaffected by any possible angular anisotropies.

The reaction product ${}^{51}$Cr decays to the first excited state in ${}^{51}$V by electron capture and subsequent emission of a characteristic E$_{\gamma} = 320$~keV photon with a half-life of T$_{1/2} = 27.7$ d and a branching probability of 9.9 \% 
\cite{Xiaolong:2006}. The activity and therefore the number of produced ${}^{51}$Cr nuclei was determined by counting the emitted $\gamma$-rays using a Pb-Cu shielded counting station at Notre Dame.
In this way the neutron detection efficiency of the detector was established with high accuracy without having to rely on cross section data. The experimental uncertainty
is dominated by the efficiency determination of the Ge detector (2\%) introduced by the uncertainty in the source activity and errors due to the activation procedure. To obtain sufficient $^{51}$V activity 
it was necessary to run with large beam currents for which the count rate in the neutron detector was too high. Therefore, the number of generated neutrons per accumulated charge had to be determined during 
several short runs with reduced beam current. This number was then used to scale the higher-current activation runs. We estimate the error resulting from this procedure to be $\pm 5\%$. The resulting total error of the activation measurement is 5.4\%.

The $^{51}$V activations were carried out at proton energies between 1600~keV to 2200~keV resulting in calibration points for neutron energies between 50~keV and 650~keV.

In addition to measurements {\sc Geant4} \cite{Agostinelli:2003} and MCNP \cite{MCNP} simulations of the setup were used to determine the energy dependence of the efficiency towards
higher neutron energies that were inaccessible to direct measurements \cite{Falahat:2010}. The geometry that was used for the simulations is shown in Fig. \ref{fig:ndetector_geant} and the resulting
efficiency can be seen together with the measurements in Fig. \ref{fig:ndetector-eff}. A more detailed description of the detector and the efficiency determination
can be found in Ref. \cite{Falahat:2012}. The simulations assumed an isotropic angular distribution of the emitted neutrons. Due to the large dimensions and almost complete $4 \pi$ coverage of the detector the effects of
possible anisotropies are strongly reduced in our measurements.

\subsection{Target thickness}\label{sec:thickness}
The observed yield $Y$ of a nuclear reaction as a function of energy can be described by \cite{Iliadis:2007}:
\begin{equation}
	Y = \int_{E_0 - \Delta E}^{E_0} \int_{0}^{E} \frac{\sigma(E')}{\varepsilon(E')} f(E,E') dE' dE \; ,
	\label{eq:integral_yield}
\end{equation}
where $E_0$ and $\Delta E$ stand for the energy of the incident particle and the target thickness in terms of energy loss; $\sigma$ and $\varepsilon$ are the energy-dependent reaction cross section
and the effective stopping power of the target material. Due to the statistical nature of energy loss a second integration over a distribution $f(E,E')$ that describes
the probability for a projectile having energy $E'$ at depth $E$ in the target can be included in the yield calculation. Under the assumption of Bohr straggling, $f(E,E')$ can be
approximated as a Gaussian with an energy-dependent width \cite{Besenbacher:1980}. The target thickness $\Delta E$ and the range of the yield integration is related to the
number density of active $(^{17}$O) target atoms $n$ through $\Delta E = n \varepsilon$ \cite{Rolfs:1988}. Therefore, the target thickness needs to be known for extracting the cross section from yield data.

The relationship between the anodization voltage ($\Delta U = 5$ V was used for the production of our targets) and the thickness of the resulting Ta$_{2}$O$_{5}$ layer 
is $t = 1.9225(V + 1.4)$ nm \cite{Seah:1988}. The expected thickness of 12.3 nm translates into a nominal area density of $n = 6.9 \times 10^{16} \frac{\text{oxygen atoms}}{\text{cm}^2}$. The thicknesses
of various targets were determined experimentally from the yields of isolated resonances in the reactions $^{17}$O$(\alpha, \gamma)^{21}$Ne (E$_{\alpha}$ = 1002~keV),
$^{17}$O$(\alpha,n)^{20}$Ne (E$_{\alpha}$ = 1247 and 1841~keV) and $^{17}$O$(\alpha, n_{1})^{20}$Ne (E$_{\alpha}$ = 1841 and 2020~keV).

\begin{figure}[tb]
	\centering
		\includegraphics[width=\columnwidth]{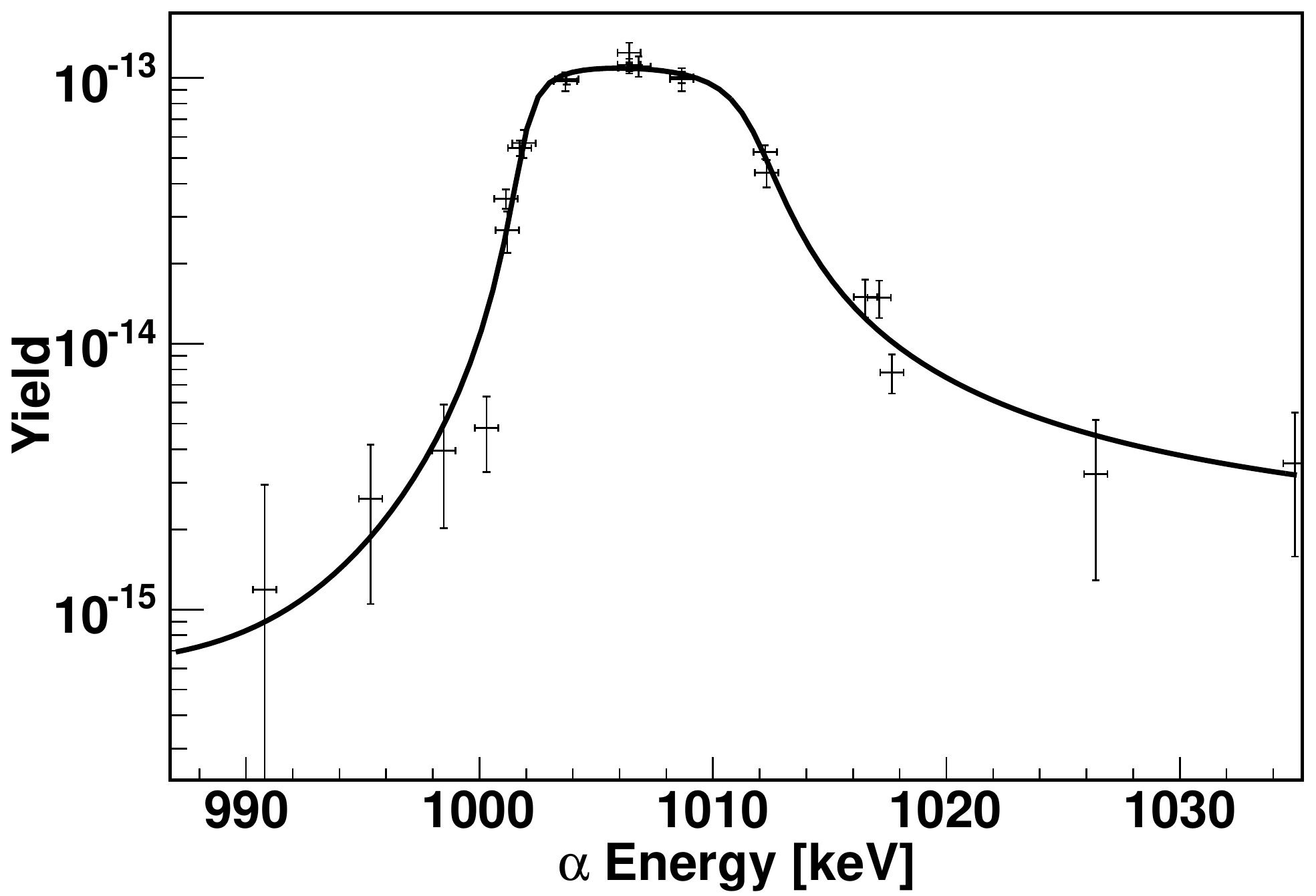}
		\caption{Yield (reactions per projectile) of the E$_{\alpha} = 1002$~keV resonance in $^{17}$O$(\alpha, \gamma)^{21}$Ne that was used for the determination of the target thickness. The line is a fit to the data points.}
	\label{fig:res-fits}
\end{figure}

Fig. \ref{fig:res-fits} shows the measured reaction yield of the E$_{\alpha} = 1002$~keV resonance in $^{17}$O$(\alpha, \gamma)^{21}$Ne. The line through the data points is a fit
of the yield based on equation \ref{eq:integral_yield}, where the single-level Breit-Wigner formula $\sigma \propto \frac{\Gamma_{\alpha} \Gamma_n}{(E-E_0)^2 + \Gamma^2/4}$ was used to describe
the energy dependence of the cross section. The average $^{17}$O area density resulting from the above mentioned 5 resonance scans is $n = (7.5 \pm 0.4) \times 10^{16}
\frac{\text{atoms}}{\text{cm}^2}$. Since a number of targets was used over the course of the experiment a slight variation in the individual target thicknesses is to be
expected. We varied the input thicknesses for the \emph{R}-matrix calculation and found the minimum $\chi^2$ value of the fits at an area density of $n =7.8 \times 10^{16}
\frac{\text{atoms}}{\text{cm}^2}$. This demonstrates that the fluctuation of the target thicknesses is within the quoted error.

\section{EXPERIMENTAL RESULTS}\label{sec:results}
\subsection{$^{17}$O$(\alpha,n_1\gamma)^{20}$Ne}\label{sec:n1_results}
\begin{figure}[tb]
	\centering
		\includegraphics[width=\columnwidth]{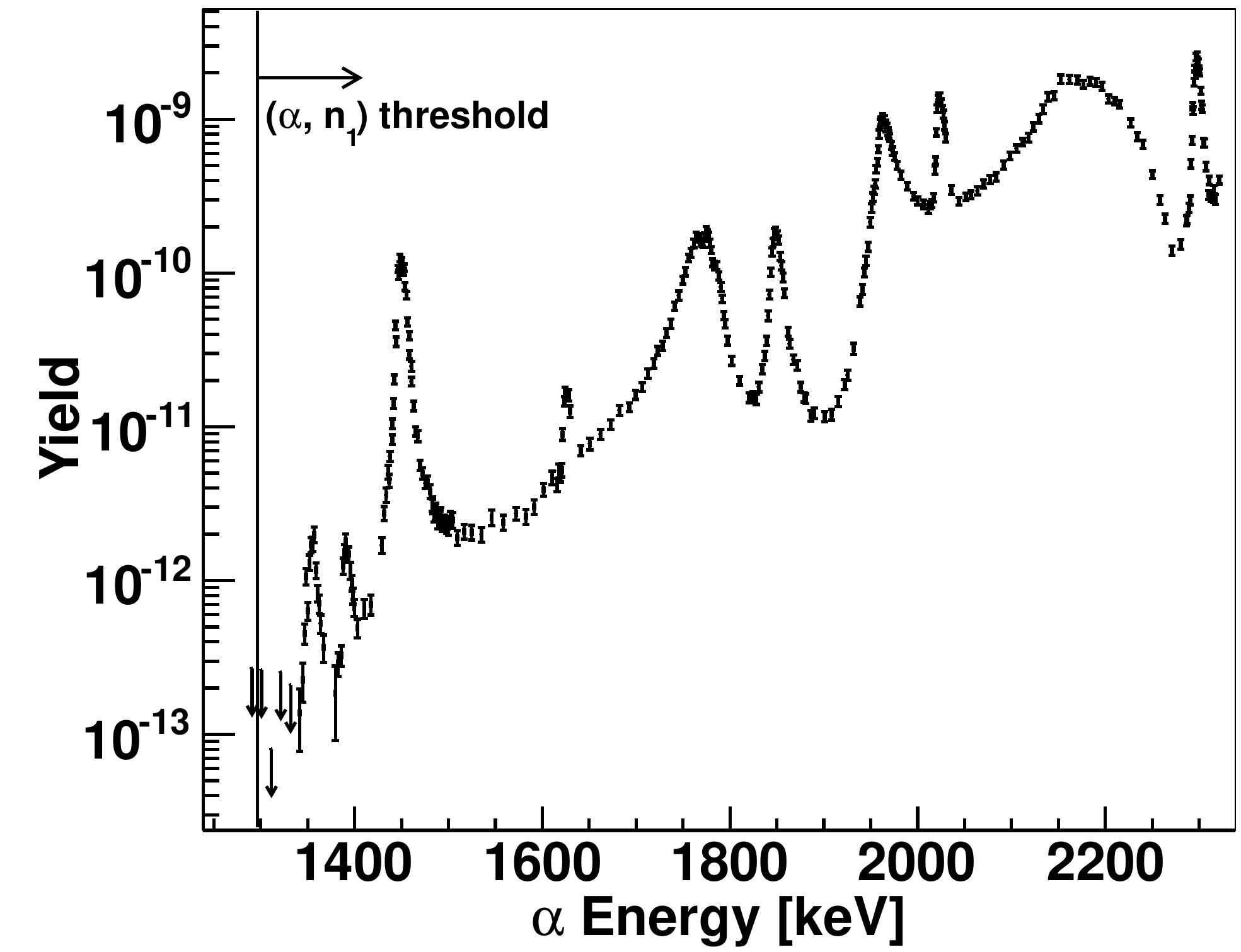}
	\caption{Yield (reactions per projectile) of the $^{17}$O$(\alpha,n_1\gamma)^{20}$Ne reaction. Also shown is the threshold of the $n_1$ channel at E$_\alpha = 1293$~keV. The arrows denote upper limits.}
	\label{fig:n1-yield}
\end{figure}
An excitation curve of the $^{17}$O$(\alpha,n_1\gamma)^{20}$Ne reaction from the $n_1$ threshold at E$_{\alpha} = 1294$~keV to 2300~keV was measured
in steps of 5~keV or less. For this measurement the E$_\gamma = 1633.7$~keV transition from the first excited state to the ground state in the $^{20}$Ne
nucleus was observed. Angular correlation effects between the direction defined by the incident beam and the detected photons were calculated to be less than 8 \% in a worst-case scenario. Since the emitted neutron is unobserved in this part of the experiment and the spin of the intermediary state is unknown angular distribution coefficients were calculated for various spin assumptions for the intermediary state in $^{21}$Ne.
The yield $Y$ (number of reactions per projectile) was calculated from the intensity $I$ in the 1633~keV peak by:
\begin{equation}
	Y = \frac{I}{Q_{dt} \eta} \; .
	\label{eq:yield}
\end{equation}
$Q_{dt} \text{ and } \eta$ represent the dead time corrected number of projectiles and the detector efficiency at 1633~keV. The resulting excitation curve is shown in Fig. \ref{fig:n1-yield}.

\subsection{$^{17}$O$(\alpha,n_{total})^{20}$Ne}\label{sec:ntot_results}
\begin{figure}[tb]
	\centering
	\includegraphics[width=\columnwidth]{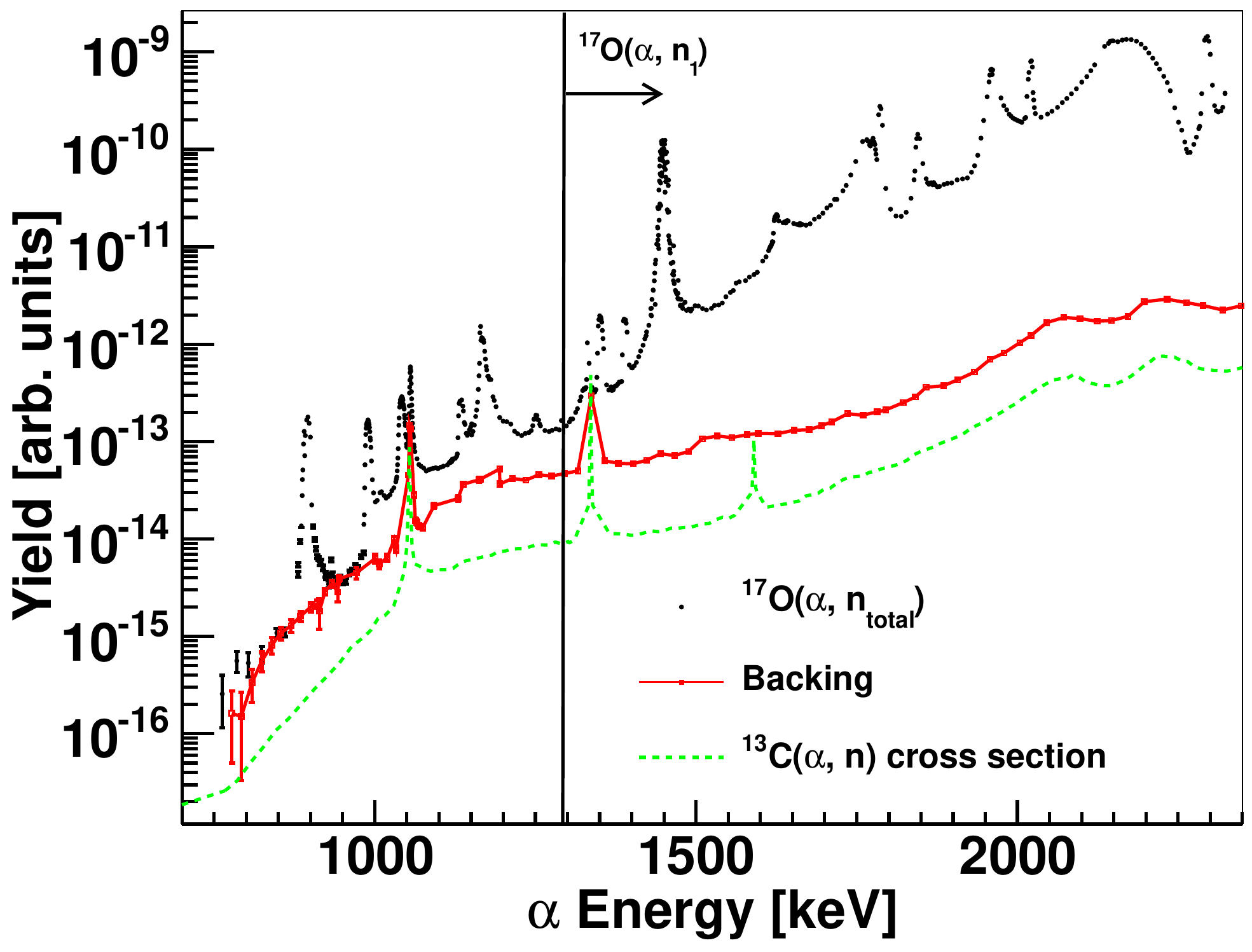}
	\caption{(Color online) Yield of the $^{17}$O$(\alpha,n_{total})^{20}$Ne reaction. Also shown in the lower curve in red is the neutron yield obtained with a blank tantalum target. The line through the blank yield is to guide the eye only. Clearly visible in both curves are the strong 1054 and 1336~keV resonances in $^{13}$C$(\alpha,n)^{16}$O. The error bars show the statistical error only. Above E$_{\alpha} = 1293$~keV both the n$_0$ and n$_1$ channels are open and contribute to the yield. The dashed line is the $^{13}$C$(\alpha,n)^{16}$O cross section from Ref. \cite{Harrisopulos:2005}. It has been scaled for comparison with the yield data.}
	\label{fig:totalyield}
\end{figure}

Fig. \ref{fig:totalyield} shows the $^{17}$O$(\alpha,n_{total})^{20}$Ne reaction yield measured with the neutron detector described in Sec. \ref{sec:n_total_setup}. The yield was not corrected for the detector efficiency.
At $\alpha$-energies above 1.3~MeV population of the first excited state in $^{20}$Ne is energetically possible and two neutron groups contribute to the yield.
The 1.05~MeV difference in energy between neutrons from the two groups, and the inability to discriminate between the two channels with the moderating detector,
results in an uncertainty in the detector efficiency. This problem will be discussed in the following sections.

Information on the initial energy of the neutron is lost during the moderation process in a detector of the type used in this part of the experiment. Therefore, background 
neutrons are in principle indistinguishable from the reaction neutrons of interest. To avoid the assignment of spurious resonances during the analysis of the data one
must carefully investigate possible beam-induced neutron producing reactions. The most important background reaction is $^{13}$C$(\alpha,n)^{16}$O due to its high cross
section and inevitable carbon buildup on beam collimators and slits. It has two strong resonances at E$_\alpha = 1054$~keV and 1336~keV \cite{Harrisopulos:2005} that lie within our energy
region of interest. To investigate the beam-induced background a yield curve was measured with a blank Ta target, covering the whole energy range of the experiment. The result of this
is shown in the lower curve in Fig. \ref{fig:totalyield}. The two $^{13}$C resonances are clearly visible in both the $^{17}$O and the blank data. The sensitivity limit has been reached towards
the lowest energy points of our measurement, and the beam-induced background dominates the yield. For comparison the $^{13}$C$(\alpha,n)^{16}$O
cross section is also shown in Fig. \ref{fig:totalyield} (scaled by a constant factor). It can be clearly seen that $^{13}$C is indeed the source of background neutrons over the
whole energy range. The typical amount of carbon can be estimated using the thin-target yield $Y = \sigma n$, where $Y$ is the backing yield and
$\sigma$ the $^{13}$C$(\alpha,n)^{16}$O cross section. The resulting $^{13}$C density is $n \approx 10^{14} \frac{\text{atoms}}{\text{cm}^2}$.
The lower energy $^{13}$C resonance was also used as an additional beam energy calibration point for our experiment.

\section{ANALYSIS}\label{sec:analysis}
\subsection{Separation of the $(\alpha, n_0)$ channel}\label{sec:a-n0}
Because of the occurrence of two neutron groups with different energies above the $n_1$ threshold at  E$_{\alpha} = 1294$~keV, the neutron detector efficiency can only be determined
accurately if the branching of the $n_0$ and $n_1$ reaction channels is known. Since we independently measured the $^{17}$O$(\alpha, n_1 \gamma)^{20}$Ne reaction by detection of the characteristic
E$_\gamma = 1633.7$~keV transition, its contribution to the $^{17}$O$(\alpha, n_{total})^{20}$Ne reaction could be subtracted from the data. As the target thicknesses used in the two experiments
were different due to the different angle of the target with respect to the beam axis ($90^{\circ}$ and $45^{\circ}$), we determined the cross section of the $^{17}$O$(\alpha, n_1 \gamma)^{20}$Ne
reaction channel with a preliminary \emph{R}-matrix fit. The simultaneous \emph{R}-matrix analysis of both channels and the computer code used for the calculations is described in
the following Section \ref{sec:r-matrix}.

\begin{figure}[tb]
	\centering
		\includegraphics[width=\columnwidth]{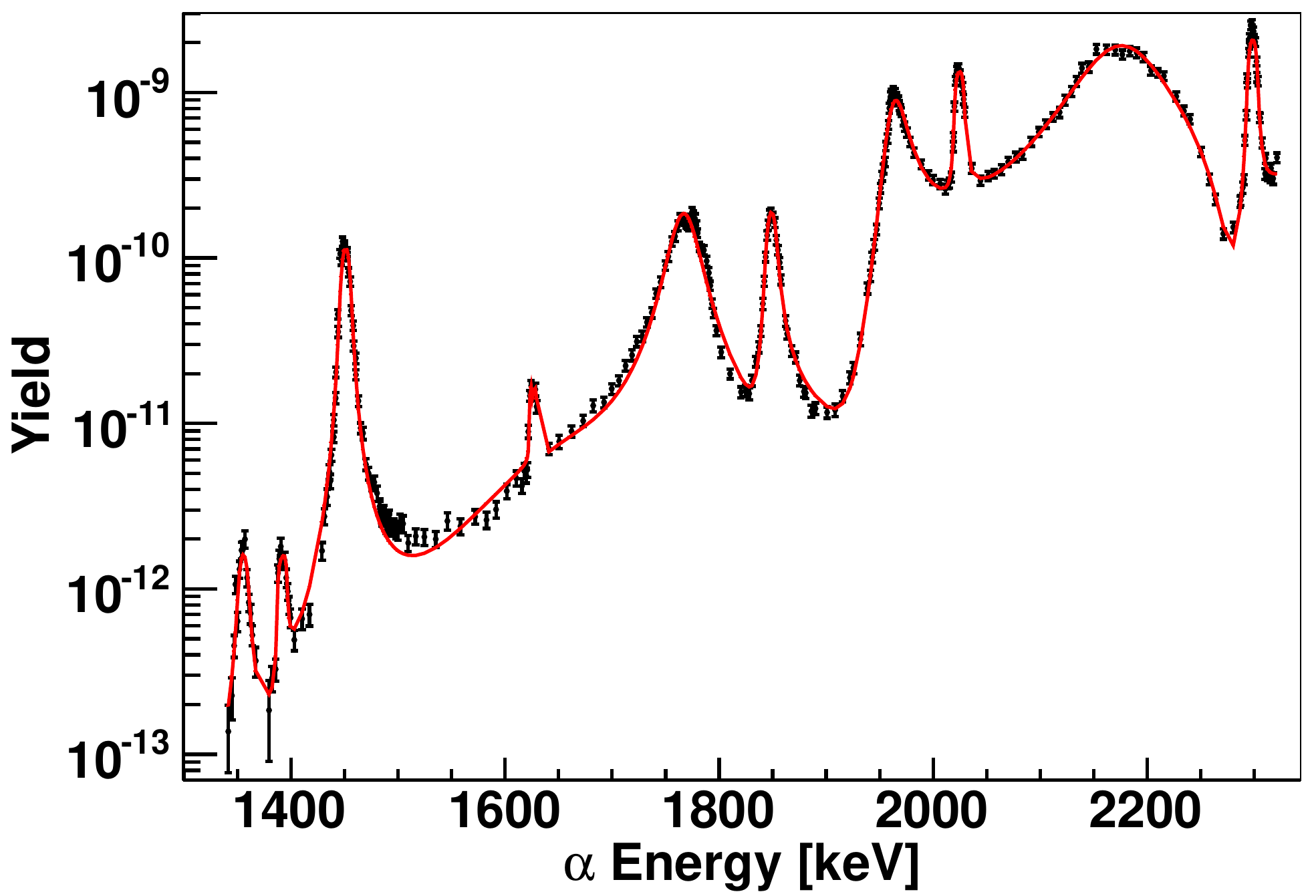}
	\caption{(Color online) Preliminary \emph{R}-matrix fit to the $^{17}$O$(\alpha,n_1 \gamma)^{20}$Ne data. The line is the fit to the data points.}
	\label{fig:n1-fit}
\end{figure}

Fig. \ref{fig:n1-fit} shows the results of the preliminary \emph{R}-matrix fit to the $(\alpha, n_1 \gamma)$ data (with $\frac{\chi^2}{\text{degrees of freedom}} = 3.0$). It was necessary to
include a background pole that destructively interferes with the E$_{\alpha} = 2150$~keV resonance. As there is little information on the J$^{\pi}$ values of the
included states available in the literature, random spins corresponding to low angular momenta were assigned. The only restriction was that states that clearly
interfered with each other had to be assigned the same spin-parity value.

The $R$ matrix cross section of the $(\alpha,n_1)$ channel $\sigma_{n_1}$ was then converted into an ``experimental'' yield contributing to the n$_{total}$ measurement (at $\alpha$ energy $E_0$) $Y_{n_1}(E_0)$ by integration of the cross
section over the target thickness $\Delta E$ and taking into account the efficiency of the neutron detector $\eta(E_{n_1})$ and the stopping power of the target material $\varepsilon(E_0)$. 90 degree neutron energies were used in this and all later calculations.

\begin{equation}
	Y_{n_1}(E_0) = \int^{E_0}_{E_0 - \Delta E} dE \frac{\sigma(E)}{\epsilon(E)} \eta(E_{n_1})
	\label{eq:n1-yield}
\end{equation}

\begin{figure}[tb]
	\centering
	\includegraphics[width=\columnwidth]{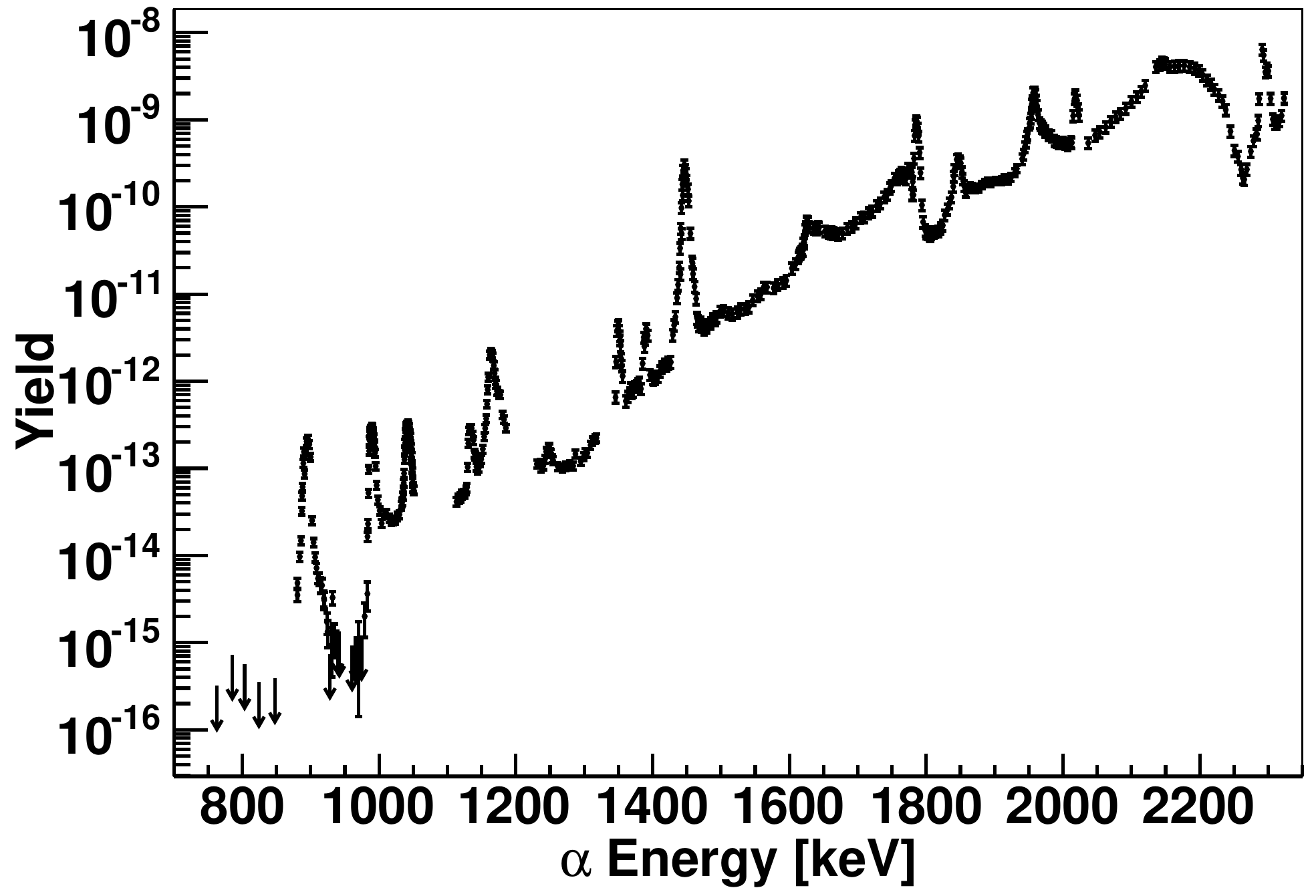}
	\caption{Yield of the $^{17}$O$(\alpha,n_0)^{20}$Ne reaction channel. For this plot the integrated $^{17}$O$(\alpha,n_1\gamma)^{20}$Ne cross section was scaled with the neutron detector efficiency and subtracted from the $^{17}$O$(\alpha,n_{total})^{20}$Ne yield. The yield in the energy regions around E$_{\alpha} = 1050$~keV and E$_{\alpha} = 1350$~keV is dominated by the two low-energy resonances in $^{13}$C$(\alpha,n)^{16}$O. These data points and the data in a region around 1200~keV where a $^{18}$O$(\alpha,n)$ resonance contributes to the yield were removed for the final analysis. The arrows denote upper limits.}
	\label{fig:n0-yield}
\end{figure}

This contribution was subtracted from the measured total neutron yield Y$_{total}$ and the result scaled with the efficiency for the neutron energy of the n$_0$ channel to give the separated n$_0$ yield Y$_{n_0}$ shown in Fig. \ref{fig:n0-yield}: $Y_{n_0}(E) = \eta(E_{n_0}) ( Y_{total}(E) - Y_{n_1}(E) )$.
Any significant effect of the neutron energy distribution on the detector efficiency cancels out by the symmetric setup about 90 degrees and by the very large angular coverage of the detector.

\subsection{\emph{R}-matrix calculations}\label{sec:r-matrix}
The \emph{R}-matrix analysis was performed in the framework of a multi-level, multi-channel approach based on the formalism outlined for the \emph{R}-matrix code AZURE \cite{Azuma:2010}.
The multi-channel capabilities of the code enabled us to simultaneously fit both measured reaction channels.  In order to directly fit the \emph{R}-matrix yield to the measured
yield data a target integration routine was added to the program. The neutron yield from the blank Ta target (see Fig. \ref{fig:totalyield} was subtracted from the n$_{total}$ data
before the final fit was conducted. The \emph{R}-matrix parameterization following Brune \cite{Brune:2002} was used in the program. The channel radii were set to 5 fm and each
target integration was divided into 25 sub-points.
It should be emphasized that because of the ambiguity in the spin assignment the
best-fit cross section can very likely be reproduced with different J$^{\pi}$ values and partial widths for the individual resonances. Therefore, the \emph{R} matrix fit should be mainly considered as
a deconvolution of the yield data and only the cross section and the resonance energies should be regarded as physically meaningful parameters.

The measured yield, a parametrisation of the stopping power of the target $\varepsilon$ (tabulated $\varepsilon$ values were calculated with the computer code SRIM-2010 \cite{Ziegler:2004}) and the
number density of ${}^{17}$O atoms in the target (as determined in Sec. \ref{sec:thickness}) were used as input for the \emph{R}-matrix calculation.  For each yield data point (at energy $E_0$) the program calculated a cross section $\sigma$, and used the
integrated yield (Eq. \ref{eq:integral_yield}) and the measured data for a least-squares fit with the Migrad algorithm from the Minuit2 minimization library \cite{Minuit2}.
It turned out that energy straggling of the projectile in the relatively thin target did not play a role and did not have to be included in the final calculation.

\begin{figure}[tb]
	\centering
	\includegraphics[width=\columnwidth]{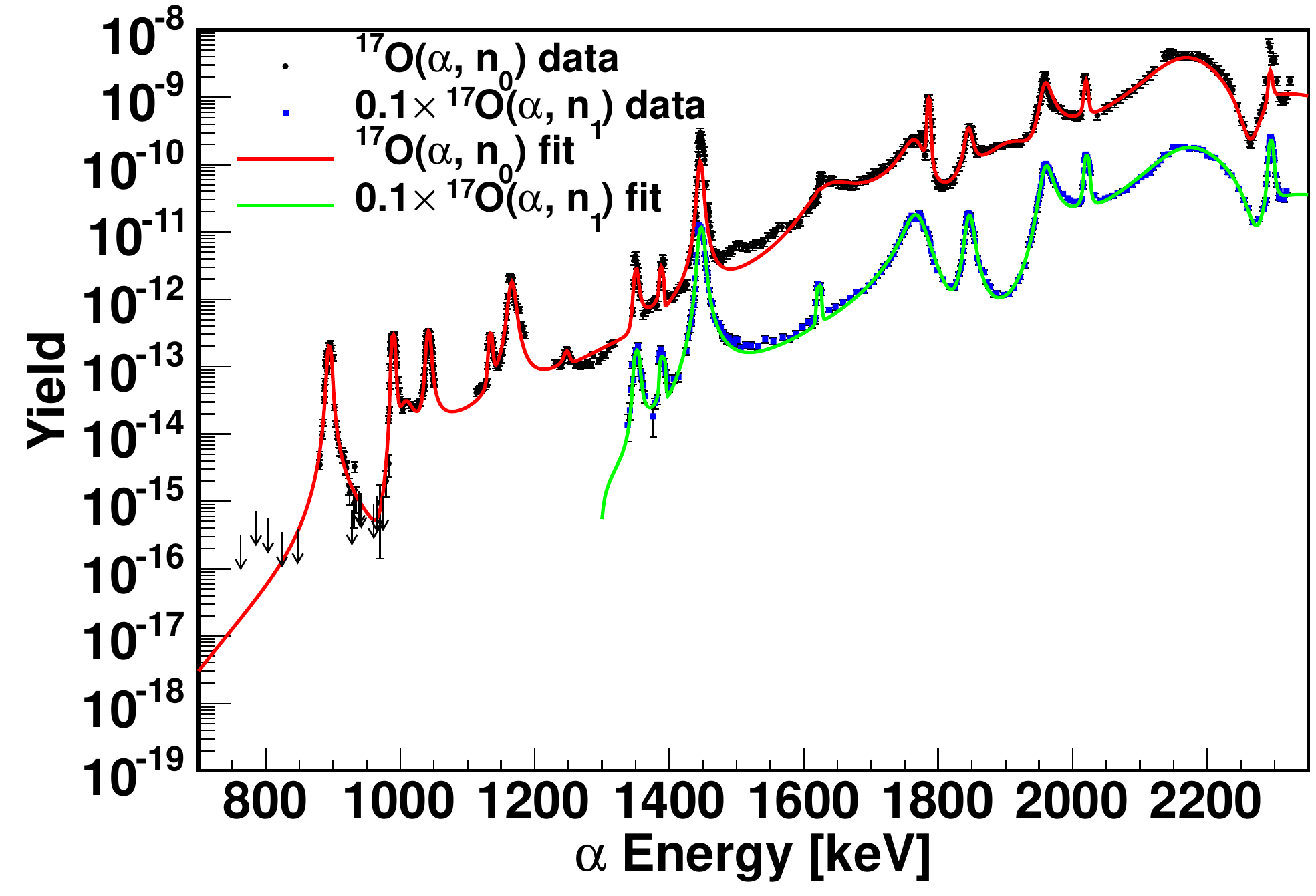}
	\caption{(Color online) Experimental and calculated yield of the $^{17}$O$(\alpha,n_0)^{20}$Ne and $^{17}$O$(\alpha,n_1 \gamma)^{20}$Ne reaction channels. The lines through the data points are the result of an \emph{R}-matrix fit to both channels. The n$_1$ channel has been divided by a factor of 10 to better separate it from the n$_0$ plot. The arrows denote upper limits.}
	\label{fig:both-n0}
\end{figure}

The same background pole as for the $n_1$ data was included in the calculation. The 1.695~MeV and 1.84~MeV resonances are only observed in the $n_0$ channel. The measured yields
and the results of the \emph{R}-matrix fit to both data sets is shown in Fig. \ref{fig:both-n0}. To make it easier to distinguish between the two channels the n$_1$ plot has
been scaled down by a factor of 10. The reduced $\chi^2$ values of the fit are 2.85 for the n$_1$ and 3.75 for the n$_0$ channel. The best-fit \emph{R}-matrix parameters from our 
calculation are shown in Table \ref{tab:parameters}. The relationship between the ``observed'' partial widths $\Gamma$ and $R$ matrix reduced widths $\gamma$ (as defined in Ref. \cite{Lane:1958})
is $\Gamma_{ic} = \frac{2 P_c \gamma_{ic}^{2}}{1 + \sum_c \gamma_{ic}^{2} \left( \frac{dS_c}{dE} \right)_{E_i}}$ \cite{Brune:2002}, where $P_C$ and $S_C$ are the penetration and the shift factor. Also
listed are the energies of previously observed states from Ref. \cite{Endt:1990}. Some of these states have been seen as resonances in $^{17}$O$(\alpha,n)$ \cite{Bair:1973} and $^{20}$Ne$(n, n)$ \cite{Mughabghab:1981}.
Known literature widths are listed for comparison. Values for the spin-parity were only available in two cases (at E$_{x} = 8.065$~MeV and E$_x = 8.68$~MeV).

\begin{figure}[tb]
	\centering
	\includegraphics[width=\columnwidth]{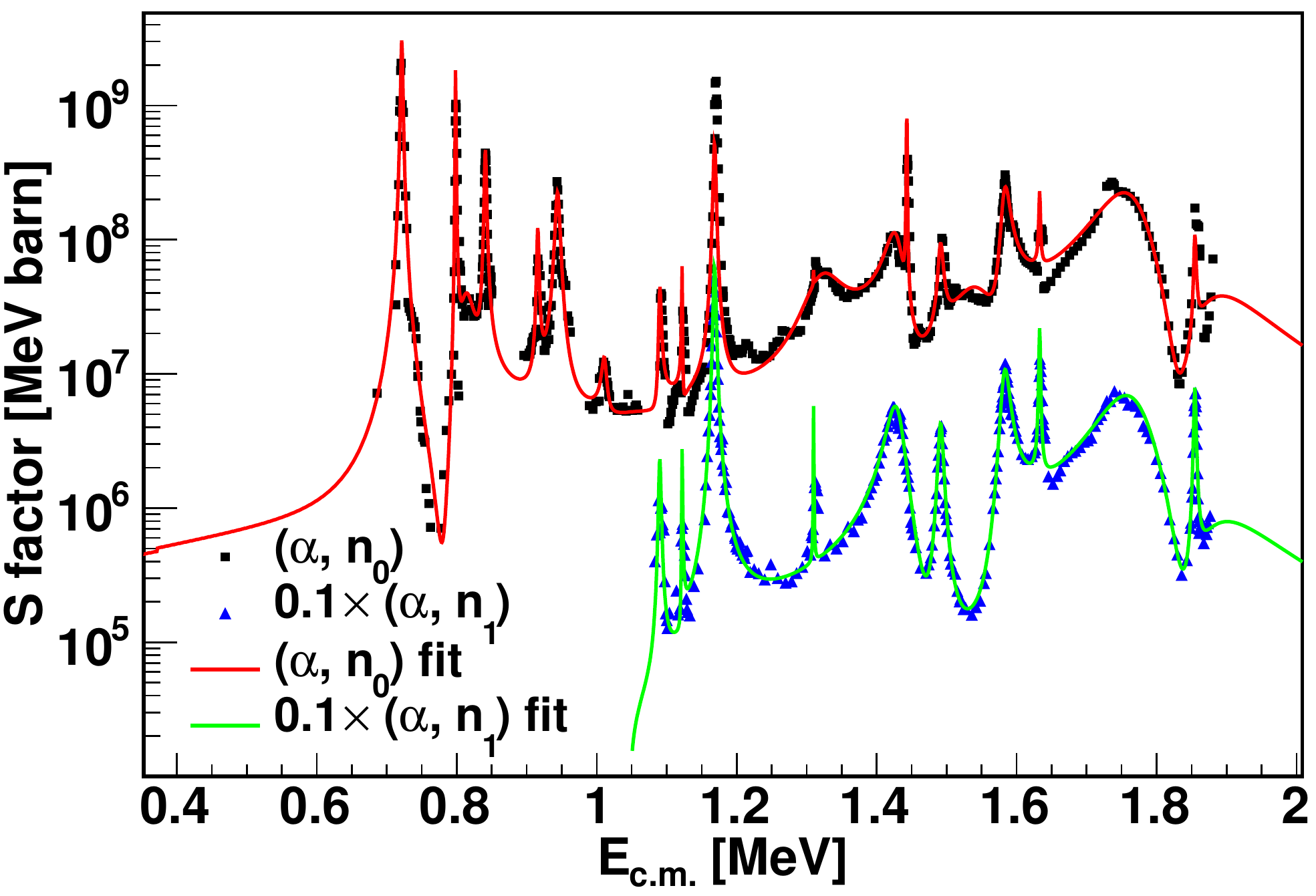}
	\caption{(Color online)  $S$-factors of the $^{17}$O$(\alpha,n_0)^{20}$Ne and $^{17}$O$(\alpha,n_1\gamma)^{20}$Ne reaction channels. Shown as the points is the deconvolved yield data (see text) for both channels and the respective \emph{R} matrix $s$-factors are represented by lines. The $s$-factors of the $(\alpha,n_1\gamma)$ channel were multiplied by a factor of 0.1 to enhance the visual separation of the curves.}
	\label{fig:cs-plots}
\end{figure}

The calculated $s$-factors for both channels are shown in Fig. \ref{fig:cs-plots}. Also shown in Fig. \ref{fig:cs-plots} are the experimental $s$-factors for both channels that
have been obtained by an least-squares deconvolution method similar to the one described in Ref. \cite{McGlone:1991}: in short, yield data was calculated by integration of
the cross section over the target thickness. The starting point was the cross section obtained from the \emph{R} matrix fit, which was then varied until the new integrated
cross section agreed with the experimental yield.

The uncertainty in the absolute scale of the $s$-factor is mostly influenced by the efficiency determination of the neutron and $\gamma$-ray detectors, and by the target thickness
that is used in the integration process. \emph{R}-matrix fits for various thicknesses within the experimentally determined range (section \ref{sec:thickness}) were performed.
The effect on the calculated $s$-factor was $\pm 10 \%$. 
Following systematic uncertainties have to be attributed to the data:
The error in the $(\alpha,n_1)$ $s$-factor was calculated by quadratic addition of the $\pm 10 \%$ from the target thickness, $\pm 8 \%$ from our estimate of the
maximal effect of the angular correlation and $\pm 5 \%$ in the efficiency determination to be $\pm 14 \%$. The uncertainty in the lower-energy $(\alpha,n_0)$ data below
the n$_1$ threshold (E$_{\alpha} < 1.3$ MeV) is $\pm 11 \%$ (10 \% target thickness and 5.5 \% detector efficiency). Finally, since the n$_0$ data above the n$_1$ threshold is the result of a
combination of both $\gamma$ and neutron measurements the associated error is the quadratic addition of both uncertainties, or $\pm 18 \%$.

\subsection{LOW-ENERGY ESTIMATE}\label{sec:extrap}
The data presented here only extend down to $E_{\alpha}^{lab} \approx 800$~keV. To cover the whole astrophysically relevant range it is necessary to estimate the contribution of lower-energy states.
13 states are known to exist between our lowest energy measurement and the $\alpha$ threshold in $^{21}$Ne \cite{Endt:1990, Firestone:2004}. Their possible contribution was evaluated as follows:
At low energies charged-particle widths are generally much smaller than neutron and $\gamma$ widths ($\Gamma_{\alpha} \ll \Gamma_{n, \gamma}$). Using an estimate of the $\alpha$ width
and the branching between the $\gamma$ and neutron channels $r = \frac{\Gamma_{\gamma}}{\Gamma_n}$ of a state the resonance strengths $\omega\gamma_{n, \gamma}$ can be evaluated from:
\begin{align}
	\omega\gamma_{(\alpha,n)} &= \omega \frac{\Gamma_{\alpha} \Gamma_n}{\Gamma_n + \Gamma_{\gamma} + \Gamma_{\alpha}} \approx \omega \Gamma_{\alpha} \frac{\Gamma_n}{\Gamma_n (r+1)} = \omega \frac{\Gamma_{\alpha}}{r + 1} \\
	\omega\gamma_{(\alpha, \gamma)} &= \omega \frac{\Gamma_{\alpha} \Gamma_{\gamma}}{\Gamma_n + \Gamma_{\gamma} + \Gamma_{\alpha}} \approx \omega \Gamma_{\alpha} \frac{\Gamma_{\gamma}}{\Gamma_{\gamma} (r^{-1}+1)} = \omega \frac{\Gamma_{\alpha}}{r^{-1} + 1}
	\label{eq:omegagamma-approx}
\end{align}

The required $\alpha$ widths can be calculated from alpha-particle widths $\Gamma_{\alpha, sp}$ and the $\alpha$ spectroscopic factor of the respective state: $\Gamma_{\alpha} = \text{C}^2 \text{S} \Gamma_{\alpha, sp}$.
The computer code {\sc DWUCK4} \cite{Kunz:2011} was used to evaluate the single-particle widths using a Woods-Saxon potential with radius $r_0 = 1.25$ fm and diffuseness
$a = 0.6$ fm. A spectroscopic factor of 0.01 was assumed for all states because no experimental information is available.

Information on the $\frac{\Gamma_{\gamma}}{\Gamma_n}$ ratios was inferred from the available literature: The relevant states have been seen in neutron scattering \cite{Mughabghab:1981},
 $^{13}$C$(^{12}$C$,\alpha)^{21}$Ne \cite{Hallock:1975} and $^{12}$C$(^{13}$C$,\alpha\gamma)^{21}$Ne \cite{Andritsopoulos:1981}, ($^{3}$He, p) \cite{Hinds:1959}
and $^{18}$O$(\alpha,n\gamma)^{21}$Ne \cite{Hoffmann:1989}. A state at $E_{x} = 7.96$~MeV has been observed in $^{16}$O$(^{7}$Li, np) \cite{Thummerer:2003}.
Levels with a known width were assumed to have a strongly suppressed $\gamma$ channel and were assigned $r = 10^{-5}$. The typical ratio between gamma and neutron widths of $10^{-3}$
was used if only the energy of the state is known. In the case of observed $\gamma$-deexcitation assignments between 0.1 and $\frac{\Gamma_{\gamma}}{\Gamma_n} = 2$ were made
based on the observed transition strengths.

Our estimates and the resulting resonance strengths of the low-energy states used in our extrapolation are listed in Tab. \ref{tab:extrap}. To obtain an upper limit on their strengths, states with unknown spin and parities
were assumed to be s-wave resonances. The lowest-lying state ($E_x = 7.36$~MeV) at only 12~keV above the $\alpha$ threshold
cannot contribute to the reaction rate and was ignored in our extrapolation.

\begin{longtable*}{cccccccc}
	\caption{$^{17}$O$+ \alpha$ single-particle widths and estimated resonance strengths. $E_x$ and $2J^{\pi}$ from Refs. \cite{Endt:1990, Firestone:2004}}\label{tab:extrap}
	\\
	\hline \hline
	$E_x$ [MeV] & $E_{cm}$ [MeV] &	$ 2J^{\pi}$ & l & $\Gamma_{sp}$ [eV]	&	$\frac{\Gamma_{\gamma}}{\Gamma_n}$	& $\omega\gamma_{\gamma}$ [$\mu \text{eV}$]	& $\omega\gamma_{n}$ [$\mu \text{eV}$]	\\
	\hline
	\endfirsthead
	\caption*{(\emph{Continued}.)}
	\\ \hline \hline
	$E_x$ [MeV] & $E_{cm}$ [MeV] &	$ 2J^{\pi}$ & l & $\Gamma_{sp}$ [eV]	&	$\frac{\Gamma_{\gamma}}{\Gamma_n}$	& $\omega\gamma_{\gamma}$ [$\mu \text{eV}$]	& $\omega\gamma_{n}$ [$\mu \text{eV}$]	\\ \hline
	\endhead
	\hline \hline
	\endlastfoot
	\hline
	\endfoot
	7.422	& 0.074&	$(7^{+}, 11^{-})$	&	2	& $1.7 \times 10^{-33}$ & 0.1 & $2.1 \times 10^{-30}$ & $2.1 \times 10^{-29}$	\\
	7.470	& 0.122&	$(1, 3)^{-}$	&	1	& $1.5 \times 10^{-22}$ & $10^{-3}$ & $5.1 \times 10^{-22}$ & $5.1 \times 10^{-19}$	\\
	7.556	& 0.208&	&	0	&	$6.0 \times 10^{-14}$	 & $10^{-3}$ & $6.0 \times 10^{-13}$ &	$6.0 \times 10^{-10}$ \\
	7.600	& 0.252&	&	0	&	$1.8 \times 10^{-11}$ & 0.1 & $1.6 \times 10^{-8}$ & $1.6 \times 10^{-7}$\\
	7.627	& 0.279&	$3^{-}$	&	1	&	$1.4 \times 10^{-10}$ & $10^{-5}$ & $9.5 \times 10^{-12}$ & $9.5 \times 10^{-7}$	\\
	7.653	& 0.305&	$7^{+} (5^{+})$	&	0	&	$3.0 \times 10^{-9}$ & 2 & $4.0 \times 10^{-5}$ & $1.3 \times 10^{-5}$	\\
	7.750	& 0.402&	&	0	& $2.2 \times 10^{-6}$ & $10^{-3}$ & $2.2 \times 10^{-5}$ & $2.2 \times 10^{-2}$\\
	7.815	& 0.467&	&	0	&	$5.4 \times 10^{-5}$ & $10^{-3}$ & $5.4 \times 10^{-4}$ & $0.54$	\\
	7.96	& 0.612&	$11^{-}$&	3	& $1.2 \times 10^{-5}$ & 2 & 0.24 & 0.08\\
	7.979	& 0.631&	$3^{-}$	&	1	&	$8.3 \times 10^{-3}$ & $10^{-5}$ & $5.5 \times 10^{-4}$ & 55\\
	7.981	& 0.633&	$(7, 11)^{+}$	&	2	&	$1.3 \times 10^{-3}$ & 1 & 8.9 & 8.9\\
	8.008	& 0.66&	$1^{-}$	&	3	&	$3.1 \times 10^{-4}$ & $10^{-5}$ & $1.0 \times 10^{-5}$ & 1.0
\end{longtable*}

The $(\alpha,n))$ reaction rate (in units of cm$^3$mol$^{-1}$s$^{-1}$) was calculated from the cross sections using the expression \cite{Fowler:1975}:

\begin{equation}
	N_A \langle \sigma v \rangle_{01} = \frac{2.074 \times 10^{10}}{T^{3/2}} \int_0^\infty E \sigma(E) e^{-11.605 E/T} dE \; .
	\label{eq:rrate-integral}
\end{equation}

The $^{17}$O$(\alpha,\gamma)^{21}$Ne rate was adopted from Ref.~\cite{Best:2011a}. Both the $(\alpha,\gamma)$ and $(\alpha,n)$ rates were extrapolated to lower temperatures using the equation for isolated, narrow resonances:
\begin{equation}
	N_A \langle \sigma v \rangle = \frac{2.643 \times 10^{10}}{T^{3/2}} \sum_i (\omega \gamma)_i e^{-11.605 E_i/T},
	\label{rrate-resonance}
\end{equation}
where resonance strengths $\omega\gamma$ were taken from Tab. \ref{tab:extrap}.

\begin{figure}[tb]
	\centering
	\includegraphics[width=1.3\columnwidth,angle=-90]{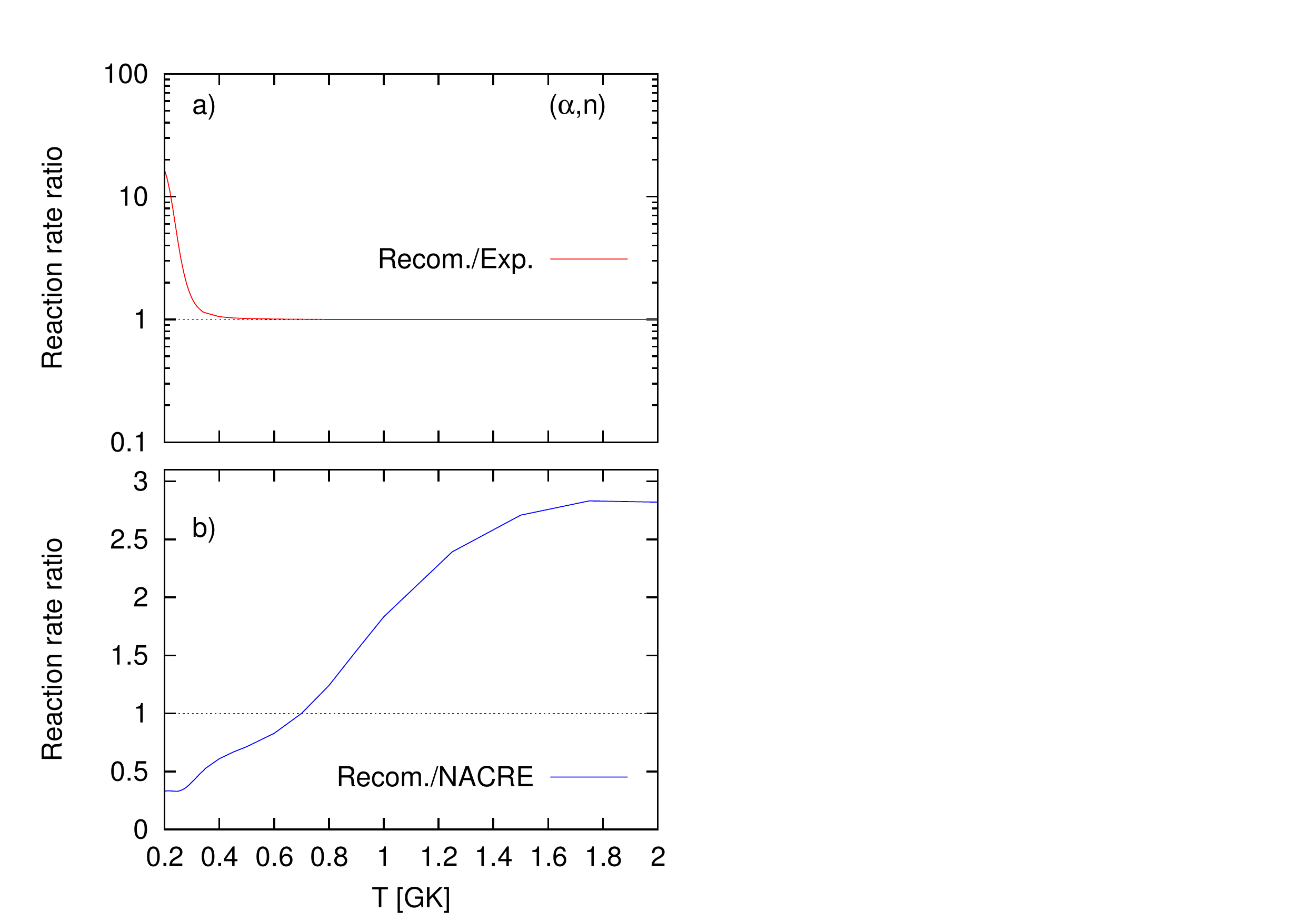}
	\caption{(Color online) Upper plot (a): Ratio of the recommended $^{17}$O$(\alpha,n)$ rate to the experimental rate. Lower plot (b): Ratio of recommended $^{17}$O$(\alpha,n)$ rate to NACRE.}
	\label{fig:n-rate}
\end{figure}

\begin{figure}[tb]
	\centering
	\includegraphics[width=1.3\columnwidth,angle=-90]{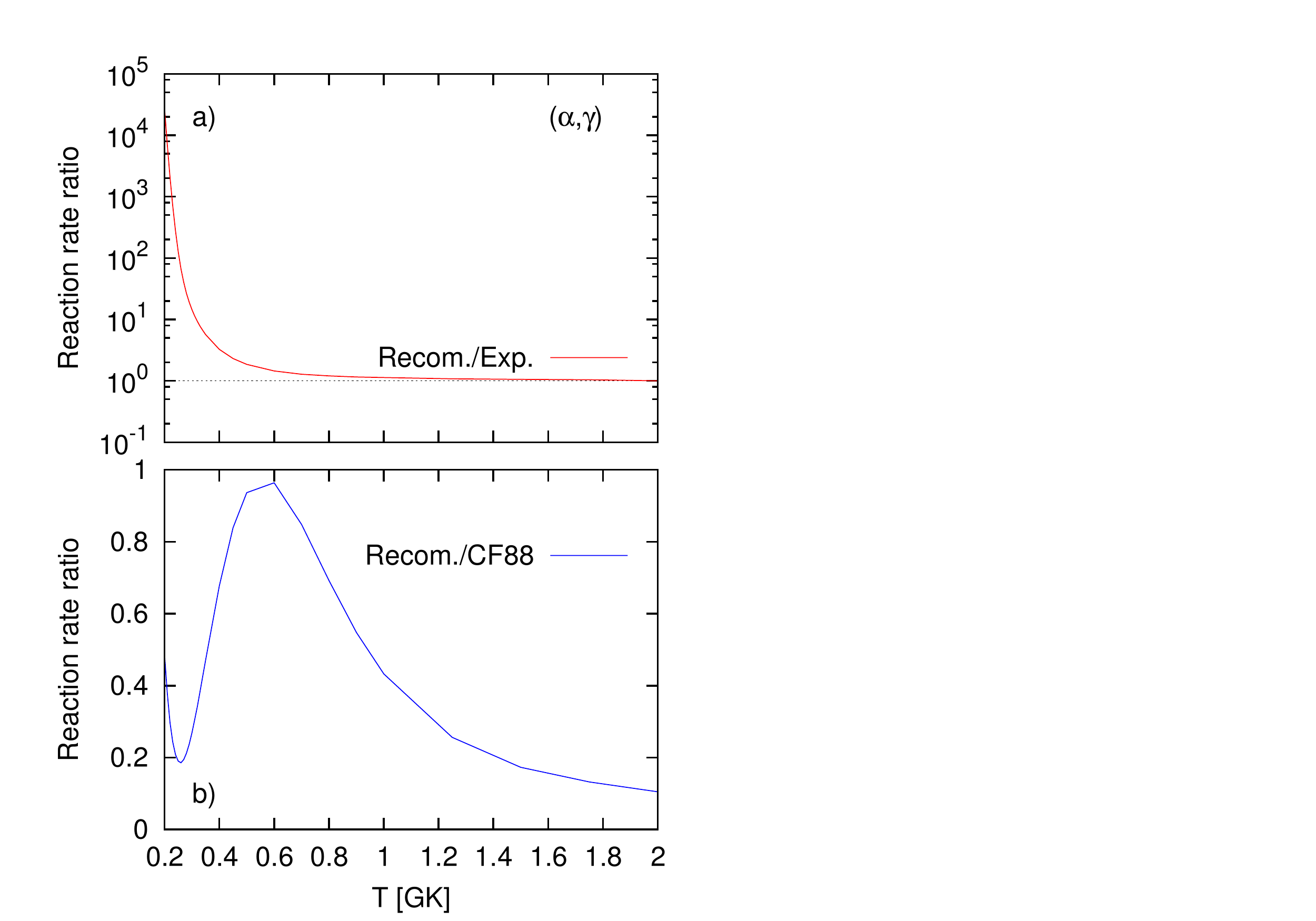}
	\caption{(Color online) Upper plot (a): Ratio of the recommended $^{17}$O$(\alpha,\gamma)$ rate to the experimental rate. Lower plot (b): Ratio of the recommended $^{17}$O$(\alpha,\gamma)$ rate to CF88.}
	\label{fig:g-rate}
\end{figure}

For both equations the temperature $T$ is given in GK and the c.m. energies $E_i$ and resonance strengths $(\omega \gamma)_i$ in~MeV.
The reaction rates are listed in Tab. \ref{tab:rates}. The ratio of our recommended $(\alpha,n)$ and $(\alpha,\gamma)$ rates to our experimental rates is shown in the upper half of Fig.~\ref{fig:n-rate} and~\ref{fig:g-rate} respectively. Shown in the lower half of the figures is a comparison of the recommended rates to literature rates. In the case of the $(\alpha,n)$ channel, the rate is shown in comparison to the more recent NACRE compilation which is based on unpublished experimental low energy data, and supersedes the previous CF88 rate. For the $(\alpha,\gamma)$ channel our rate is compared to the CF88 rate because NACRE does not list a rate for this reaction.

\begin{figure}[!tb]
	\centering
	\includegraphics[width=0.7\columnwidth,angle=-90]{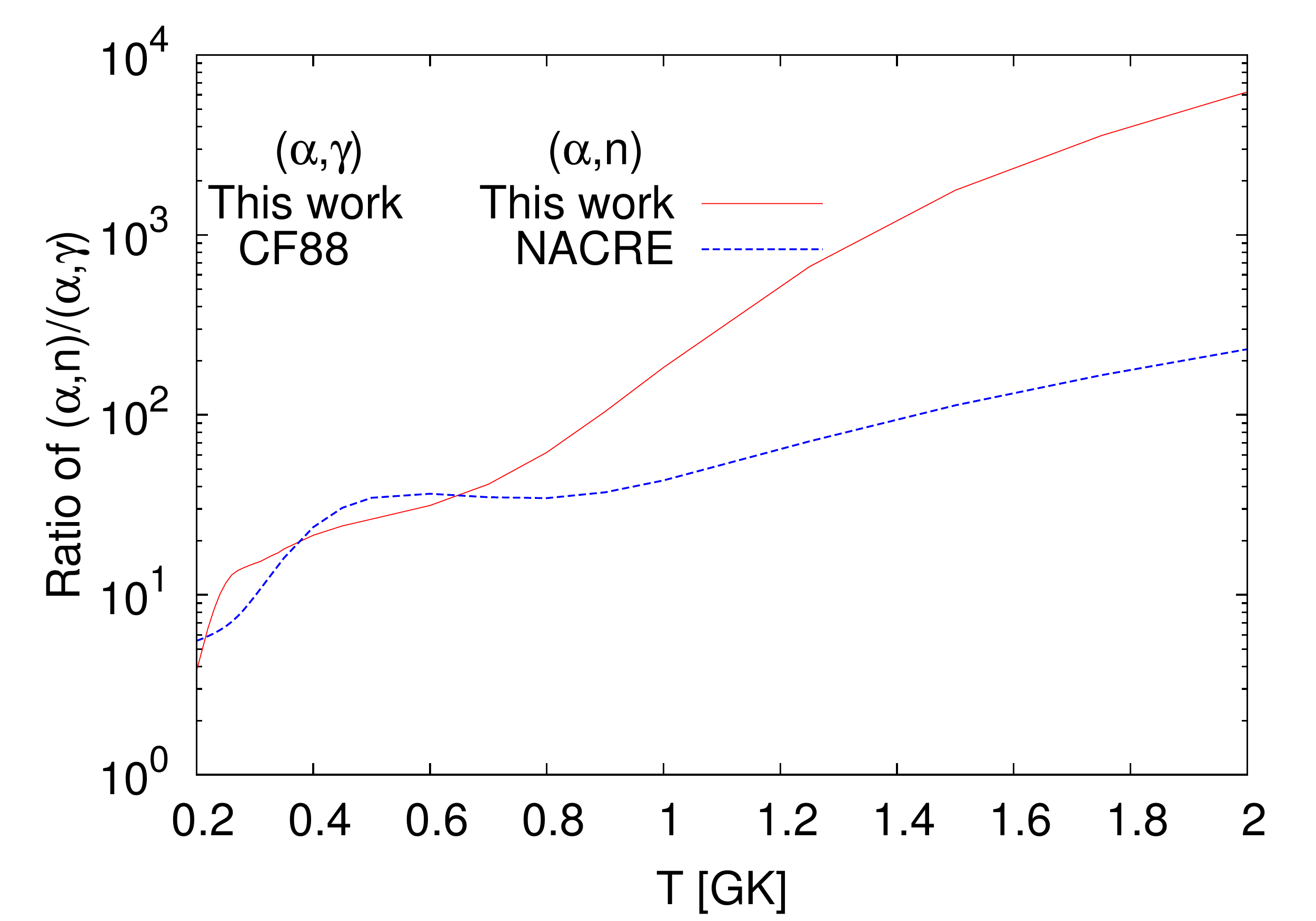}
	\caption{(Color online) The $(\alpha,n)/(\alpha,\gamma)$ ratio as a function of temperature using the recommended rates, and compilations.}
	\label{fig:an-ag-ratio}
\end{figure}

Plotted in Fig.~\ref{fig:an-ag-ratio} is the ratio of the NACRE $(\alpha,n)$ to the CF88 $(\alpha,\gamma)$ rate, as well as the same ratio using the recommended rates from this work. At the temperatures of relevance for the $s$ process, these two ratios are within a factor of 2. 

\subsection{Astrophysical implications}\label{sec:astro}
As discussed in the introduction, the $^{17}$O($\alpha$,$n$)/$^{17}$O($\alpha$,$\gamma$) reaction rate ratio determines how efficiently $^{16}$O acts as a neutron poison for the $s$ process in massive stars.
Uncertainty in this ratio predominantly arose from a lack of knowledge regarding the strength of the ($\alpha$,$\gamma$) channel. In previous estimations the ratio ranges from between $\approx 10$ (if using
the CF88 rates \cite{CF88}) to $\approx 10000$ (if~\cite{Descouvemont:1993} is used).

In the present work, both the $(\alpha,n)$ and the $(\alpha,\gamma)$ rates are weaker than those proposed by NACRE and CF88, respectively, at He burning temperatures. For instance, at $T$ $\sim$ 0.3~GK, the recommended $(\alpha,n)$ rate is about $1/4$ of NACRE, and the $(\alpha,\gamma)$ rate is about $1/4$ of the CF88 rate. 

\begin{figure*}[tb]
\centering
\includegraphics[width=0.5\textwidth,angle=270]{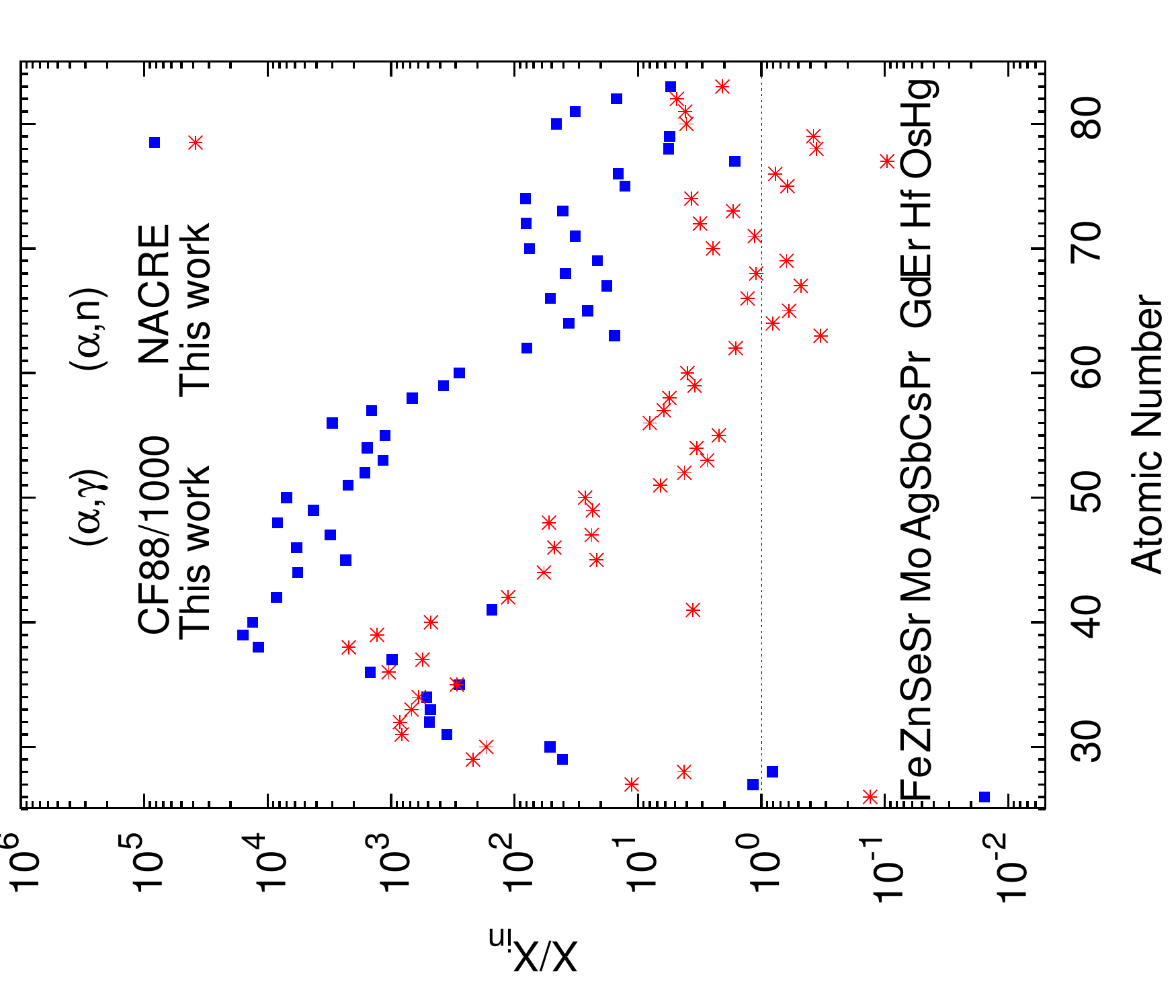}
\caption{(Color online)
Comparison of $s$ process elemental abundances obtained at the end of convective He-core burning, relative to the initial abundance distribution. 
Red stars are the $s$ process abundances using the $^{17}$O($\alpha,\gamma)^{21}$Ne and $^{17}$O($\alpha,n)^{20}$Ne rates obtained in this work
(where the ratio between the two rates is $\approx 10$ at T=0.3 GK). Blue squares are abundances obtained using the 
$^{17}$O($\alpha$,$\gamma)^{21}$Ne CF88 rate as modified by Descouvemont et al.~\cite{Descouvemont:1993}, and the $^{17}$O($\alpha,n)^{20}$Ne rate from NACRE (where the ratio between the rates is $\approx 10000$ at T=0.3 GK). 
See text for details.}    
\label{fig:abund}
\end{figure*}

Shown in Fig.~\ref{fig:abund} is the effect of our recommended $^{17}$O($\alpha$,$\gamma$) and $^{17}$O($\alpha$,$n$) rates (red stars) on $s$ process elemental abundances at the end of He core burning for fast rotating massive stars.  This is compared to the abundances obtained using the $^{17}$O$(\alpha,n)$ rate adopted from NACRE~\cite{Angulo:1999} and the $^{17}$O($\alpha$,$\gamma$) rate from CF88, modified by Descouvemont (blue squares). In both cases, the $s$ process elemental abundances are plotted relative to the initial abundance. For comparison to Fig.~1 of Ref.~\cite{Frischknecht:2012}, Fig.~\ref{fig:abund} is plotted as a function of atomic number. We also performed the calculations using the $(\alpha,\gamma)$ rate of CF88 and the $(\alpha,n)$ rate of NACRE. The resulting elemental abundances are very close to the results obtained with the present rates, because both sets of rates have similar $(\alpha,n)/(\alpha,\gamma)$ ratios (see Fig.~\ref{fig:an-ag-ratio}).

Nucleosynthesis calculations were performed using the post-processing network code PPN~\cite{Hirschi:2008}. 
The single-zone trajectory was extracted from a complete 25 solar mass star (e.g., Hirschi~\cite{Hirschi:2007}), calculated using the Geneva stellar evolution code (GENEC) which is described in detail in Refs.~\cite{Hirschi:2004} and~\cite{Eggenberger:2008}, and assuming an initial metallicity of Z = 10$^{-5}$. In order to take into account the primary $^{22}$Ne production arising from stellar rotational mixing~\cite{Pignatari:2008,Frischknecht:2012}, the initial abundance of $^{22}$Ne was set to be 1\%, in accordance with Hirschi et al.~\cite{Hirschi:2008a}. 

Figure~\ref{fig:abund} demonstrates the sensitivity of the $s$ process in low-metallicity stars to the ratio of the $(\alpha,n)/(\alpha,\gamma)$ channels. Our new measurement of the ratio confirms that the $s$ process is not suppressed in fast-rotating massive stars, and that the main region receiving the additional feeding is between the Fe seeds up to Ba, peaking with maximum efficiency in the Sr region. Compared to the Descouvemont-modified CF88 $^{17}$O$(\alpha,\gamma)$ ratio however, we observe in general less $s$ process efficiency because the modified $(\alpha,\gamma)$ rate is many orders of magnitude smaller than our measurement. As a result the neutron recycling is far more efficient. 

\section{CONCLUSION}\label{sec:conclusions}
We have presented new experimental results and \emph{R}-matrix calculations for the $^{17}$O($\alpha,n)^{20}$Ne and the $^{17}$O($\alpha, n_1 \gamma)^{20}$Ne reactions. A reaction rate 
for the $(\alpha,n)$ and the competing $(\alpha,\gamma)$ channel has been calculated and extended to lower temperatures by estimates on the contributions of resonances in lower-lying 
states. Despite having lower $(\alpha,n)$ and $(\alpha,\gamma)$ rates than NACRE and CF88, the new recommended ratio has a similar impact on the $s$ process elemental abundances.  This is because the difference between the two ratios is less than a factor of 2 in the temperature range of relevance (see Fig.~\ref{fig:an-ag-ratio}). Such minor changes would not be apparent on a log scale as used in Fig.~\ref{fig:abund}. Using the CF88 $(\alpha,\gamma)$ and the NACRE $(\alpha,n)$ rates for a star with an average equatorial velocity of $\langle v_{eq} \rangle=428$~km~s$^{-1}$, Ref.~\cite{Frischknecht:2012} finds an $s$ process elemental distribution which is similar to this work. Therefore, we confirm that massive rotating stars do play a significant role in the production of elements up to the Sr neutron magic peak. 

However significant uncertainties still arise from unobserved low energy resonances. A calculation to test the impact of the low energy resonances strengths shows that the nucleosynthesis is not sensitive to the absolute magnitude of the resonances, but rather to the relative ratio of the ($\alpha,\gamma$)/($\alpha$,n) reaction channels. Changing the ratio in the favor of ($\alpha$,n) by a factor of 3 results in significant enhancement of the A~=~40 -- 60 region.

These results indicate that $^{16}$O is a significant neutron poison in metal poor massive rotating stars. An extension of experimental data towards lower energies can further improve the precision
of the reaction rates and seems feasible.

\acknowledgments{The authors express their gratitude to the technical staff of the Nuclear Science Laboratory at Notre Dame. 
This work was funded by the National Science Foundation through grant number Phys-0758100 and the Joint Institute for Nuclear Astrophysics supported through the NSF Physics
Frontier Center program, grant number Phys-0822648. M.P. acknowledges support from the Ambizione grant of the SNSF (MP, Switzerland), EU MIRG-CT-2006-046520, and EuroGenesis (MASHE). M.B. acknowledges support from the Alliance Program of the Helmholtz Association (AH216/EMMI).}

\begin{longtable}{c c c c c c c c}
	\caption{Best-fit \emph{R}-matrix parameters for the fits shown in figures \ref{fig:both-n0} and \ref{fig:cs-plots}. l and s are the relative angular momentum and the channel spin and $\Gamma$ is the ``observed'' partial width as defined in Brune \cite{Brune:2002}. The ``$+/-$'' column shows the relative interference signs needed to reproduce our result. The states are sorted by ascending total angular momentum J.}\label{tab:parameters}
\\ \hline \hline E$_x$ [MeV] & J$^\pi$ & Channel & l & s & $\Gamma$ [eV] & +/- & E$_x$ (lit.) \cite{Endt:1990} \\ \hline
\endfirsthead
\caption*{(\emph{Continued}.)}
\\ \hline \hline E$_x$ [MeV] & J$^\pi$ & Channel & l & s & $\Gamma$ [eV] & +/- & E$_x$ (lit.) \cite{Endt:1990} \\ \hline
\endhead
\hline \hline
\endlastfoot
\hline
\endfoot
8.791 & 1/2$^{+}$ & $\alpha$ & 2 & 5/2 & 627 & + &\\*
 & & n$_0$ & 0 & 1/2 & 342 & + & \\
8.189 & 3/2$^{-}$ & $\alpha$ & 1 & 5/2 & 88.4 $\times 10^{-3}$ & + & 8.186 (10) \\*
 & & n$_0$ & 1 & 1/2 & 3.32 $\times 10^3$ & - & \\
8.292 & 3/2$^{-}$ & $\alpha$ & 1 & 5/2 & 559 $\times 10^{-3}$ & + & 8.287 (10) \\*
 & & n$_0$ & 1 & 1/2 & 7.54 $\times 10^3$ & + & \\
8.438 & 3/2$^{-}$ & $\alpha$ & 1 & 5/2 & 563 $\times 10^{-3}$ & + & 8.436 (10) \\*
 & & n$_1$ & 1 & 3/2 & 1.19 $\times 10^3$ & + & \\*
 & & & 1 & 5/2 & 140 & + & \\*
 & & n$_0$ & 1 & 1/2 & 2.34 $\times 10^3$ & + & \\
8.47 & 3/2$^{-}$ & $\alpha$ & 1 & 5/2 & 363 $\times 10^{-3}$ & + & 8.470 (10) \\*
 & & n$_1$ & 1 & 3/2 & 320 & + & \\*
 & & & 1 & 5/2 & 37.7 & + & \\*
 & & n$_0$ & 1 & 1/2 & 781 & - & \\
8.665 & 3/2$^{-}$ & $\alpha$ & 1 & 5/2 & 76.1 & - & 8.680 (7) \\*
 & & n$_0$ & 1 & 1/2 & 62.5 $\times 10^3$ & + & J$^{\pi} = 3/2^-$ \\
 & &       &   &	&  &  & $\Gamma = 54 (6)$ keV \\
8.899 & 3/2$^{-}$ & $\alpha$ & 1 & 5/2 & 531 & + & \\*
 & & n$_0$ & 1 & 1/2 & 75.2 $\times 10^3$ & + & \\
9.203 & 3/2$^{-}$ & $\alpha$ & 1 & 5/2 & 1.83 $\times 10^3$ & + & \\*
 & & n$_1$ & 1 & 3/2 & 758 & + & \\*
 & & & 1 & 5/2 & 85.6 & + & \\*
 & & n$_0$ & 1 & 1/2 & 1.03 $\times 10^3$ & + & \\
8.069 & 3/2$^{+}$ & $\alpha$ & 2 & 5/2 & 46.2 $\times 10^{-3}$ & + & 8.065 (10) \\*
 & & n$_0$ & 2 & 1/2 & 3.0 $\times 10^3$ & + &   J$^{\pi} = 3/2^+$ \\*
  & &       &   &   & &  & $\Gamma = 8 (3)$ keV \\
8.146 & 3/2$^{+}$ & $\alpha$ & 2 & 5/2 & 54.7 $\times 10^{-3}$ & + & \\*
 & & n$_0$ & 2 & 1/2 & 974 & + & \\
8.359 & 3/2$^{+}$ & $\alpha$ & 2 & 5/2 & 58.3 $\times 10^{-3}$ & + & \\*
 & & n$_0$ & 2 & 1/2 & 7.99 $\times 10^3$ & - & \\
8.839 & 3/2$^{+}$ & $\alpha$ & 2 & 5/2 & 131 & + & 8.849 (5) \\*
 & & n$_1$ & 0 & 3/2 & 786 & + & $\Gamma = 10$ keV \\*
 & & & 2 & 3/2 & 84.1 & + & \\*
 & & & 2 & 5/2 & 2.18 $\times 10^3$ & + & \\*
 & & n$_0$ & 2 & 1/2 & 5.13 $\times 10^3$ & + & \\
8.981 & 3/2$^{+}$ & $\alpha$ & 2 & 5/2 & 403 & + & \\*
 & & n$_1$ & 0 & 3/2 & 19.4 & + & \\*
 & & & 2 & 3/2 & 1.77 $\times 10^{-3}$ & + & \\*
 & & & 2 & 5/2 & 1.06 $\times 10^3$ & + & \\*
 & & n$_0$ & 2 & 1/2 & 881 & - & \\
8.264 & 5/2$^{-}$ & $\alpha$ & 1 & 5/2 & 54.4 $\times 10^{-3}$ & + & \\*
 & & n$_0$ & 3 & 1/2 & 3.3 $\times 10^3$ & + & \\
8.516 & 5/2$^{-}$ & $\alpha$ & 1 & 5/2 & 21.3 & - & 8.526 (2) \cite{Bair:1973} \\*
 & & n$_1$ & 1 & 3/2 & 1.07 $\times 10^3$ & - & $\Gamma = 6$ keV \\*
 & & & 1 & 5/2 & 1.22 $\times 10^3$ & - & \\*
 & & n$_0$ & 3 & 1/2 & 1.86 $\times 10^3$ & + & \\
8.16 & 5/2$^{+}$ & $\alpha$ & 0 & 5/2 & 16.0 $\times 10^{-3}$ & - & \\*
 & & & 2 & 5/2 & 0.45 $\times 10^{-3}$ & + & \\*
 & & n$_0$ & 2 & 1/2 & 23.0 $\times 10^3$ & + & \\
8.774 & 5/2$^{+}$ & $\alpha$ & 0 & 5/2 & 217 & - & \\*
 & & & 2 & 5/2 & 32.4 & + & \\*
 & & n$_1$ & 2 & 3/2 & 448 & + & \\*
 & & & 0 & 5/2 & 10.3 $\times 10^3$ & + & \\*
 & & & 2 & 5/2 & 9.35 & + & \\*
 & & n$_0$ & 2 & 1/2 & 18.7 $\times 10^3$ & - & \\
8.929 & 5/2$^{+}$ & $\alpha$ & 0 & 5/2 & 31.5 & - & 8.930 (5) \\*
 & & & 2 & 5/2 & 768 & + & $\Gamma = 5$ keV \\*
 & & n$_1$ & 2 & 3/2 & 1.03 & + & \\*
 & & & 0 & 5/2 & 4.43 $\times 10^3$ & + & \\*
 & & & 2 & 5/2 & 24.9 $\times 10^{-3}$ & + & \\*
 & & n$_0$ & 2 & 1/2 & 9.34 $\times 10^3$ & - & \\
9.099 & 5/2$^{+}$ & $\alpha$ & 0 & 5/2 & 16.8 $\times 10^3$ & + & \\*
 & & & 2 & 5/2 & 1.1 $\times 10^3$ & + & \\*
 & & n$_1$ & 2 & 3/2 & 37.7 & - & \\*
 & & & 0 & 5/2 & 16.7 $\times 10^3$ & + & \\*
 & & & 2 & 5/2 & 816 $\times 10^{-3}$ & + & \\*
 & & n$_0$ & 2 & 1/2 & 55.0 $\times 10^3$ & + & \\
9.232 & 5/2$^{+}$ & $\alpha$ & 0 & 5/2 & 4.69 $\times 10^3$ & + & \\*
 & & & 2 & 5/2 & 2.92 $\times 10^3$ & + & \\*
 & & n$_1$ & 2 & 3/2 & 20.7 & - & \\*
 & & & 0 & 5/2 & 20.8 $\times 10^3$ & + & \\*
 & & & 2 & 5/2 & 444 $\times 10^{-3}$ & + & \\*
 & & n$_0$ & 2 & 1/2 & 110 $\times 10^3$ & + & \\
8.658 & 9/2$^{-}$ & $\alpha$ & 3 & 5/2 & 625 $\times 10^{-3}$ & + & \\*
 & & n$_1$ & 3 & 3/2 & 222 & + & \\*
 & & & 3 & 5/2 & 222 & + & \\*
 & & n$_0$ & 5 & 1/2 & 109 & + & \\
 \end{longtable}

\begin{longtable}{ccccc}
	\caption{$^{17}$O$(\alpha,n)^{20}$Ne and $^{17}$O$(\alpha,\gamma)^{21}$Ne reaction rates (in units of cm$^3$ mol$^{-1}$ s$^{-1}$). Above T = 2 GK the recommended rates are Hauser-Feshbach calculations, below they are the sum of the respective low-energy estimates and the experimental data.}\label{tab:rates}
	\\
	\hline \hline
	& \multicolumn{2}{c}{$(\alpha,n)$} & \multicolumn{2}{c}{$(\alpha,\gamma)$} \\
	T [GK] & $N_A \langle \sigma v \rangle_{exp.}$ & $N_A \langle \sigma v \rangle_{recom.}$ & $N_A \langle \sigma v \rangle_{exp.}$ & $N_A \langle \sigma v \rangle_{recom.}$ \\
	\hline
	\endfirsthead
	\caption*{(\emph{Continued}.)}
	\\ \hline \hline
	& \multicolumn{2}{c}{$(\alpha,n)$} & \multicolumn{2}{c}{$(\alpha,\gamma)$} \\
	T [GK] & $N_A \langle \sigma v \rangle_{exp.}$ & $N_A \langle \sigma v \rangle_{recom.}$ & $N_A \langle \sigma v \rangle_{exp.}$ & $N_A \langle \sigma v \rangle_{recom.}$ \\ \hline
	\endhead
	\hline \hline
	\endlastfoot
	\hline
	\endfoot
0.1  &  9.96 $\times 10^{-22}$ &  5.54 $\times 10^{-20}$  &  9.52 $\times 10^{-38}$ &  1.67 $\times 10^{-20}$ \\
0.11  &  1.44 $\times 10^{-20}$ &  6.89 $\times 10^{-19}$   &  4.24 $\times 10^{-34}$ &  3.37 $\times 10^{-19}$ \\
0.12  &  1.54 $\times 10^{-19}$ &  6.13 $\times 10^{-18}$   &  4.61 $\times 10^{-31}$ &  4.17 $\times 10^{-18}$ \\
0.13  &  1.28 $\times 10^{-18}$ &  4.21 $\times 10^{-17}$   &  1.69 $\times 10^{-28}$ &  3.50 $\times 10^{-17}$ \\
0.14  &  8.76 $\times 10^{-18}$ &  2.40 $\times 10^{-16}$   &  2.65 $\times 10^{-26}$ &  2.16 $\times 10^{-16}$ \\
0.15  &  5.03 $\times 10^{-17}$ &  1.20 $\times 10^{-15}$   &  2.10 $\times 10^{-24}$ &  1.04 $\times 10^{-15}$ \\
0.16  &  2.49 $\times 10^{-16}$ &  5.44 $\times 10^{-15}$   &  9.58 $\times 10^{-23}$ &  4.13 $\times 10^{-15}$ \\
0.17  &  1.09 $\times 10^{-15}$ &  2.25 $\times 10^{-14}$   &  2.77 $\times 10^{-21}$ &  1.38 $\times 10^{-14}$ \\
0.18  &  4.35 $\times 10^{-15}$ &  8.51 $\times 10^{-14}$   &  5.49 $\times 10^{-20}$ &  4.03 $\times 10^{-14}$ \\
0.2  &  5.64 $\times 10^{-14}$ &  9.33 $\times 10^{-13}$  &  8.69 $\times 10^{-18}$ &  2.48 $\times 10^{-13}$ \\
0.25  &  2.66 $\times 10^{-11}$ &  1.11 $\times 10^{-10}$   &  7.52 $\times 10^{-14}$ &  9.56 $\times 10^{-12}$ \\
0.3  &  4.37 $\times 10^{-9}$ &  6.50 $\times 10^{-9}$  &  3.01 $\times 10^{-11}$ &  4.35 $\times 10^{-10}$ \\
0.35  &  1.89 $\times 10^{-7}$ &  2.14 $\times 10^{-7}$   &  2.10 $\times 10^{-9}$ &  1.19 $\times 10^{-8}$ \\
0.4  &  3.28 $\times 10^{-6}$ &  3.45 $\times 10^{-6}$  &  4.94 $\times 10^{-8}$ &  1.61 $\times 10^{-7}$ \\
0.45  &  3.06 $\times 10^{-5}$ &  3.14 $\times 10^{-5}$   &  5.63 $\times 10^{-7}$ &  1.30 $\times 10^{-6}$ \\
0.5  &  1.86 $\times 10^{-4}$ &  1.89 $\times 10^{-4}$  &  3.88 $\times 10^{-6}$ &  7.17 $\times 10^{-6}$ \\
0.6  &  3.06 $\times 10^{-3}$ &  3.08 $\times 10^{-3}$  &  6.79 $\times 10^{-5}$ &  9.82 $\times 10^{-5}$ \\
0.7  &  2.67 $\times 10^{-2}$ &  2.68 $\times 10^{-2}$  &  5.07 $\times 10^{-4}$ &  6.51 $\times 10^{-4}$ \\
0.8  &  1.67 $\times 10^{-1}$ &  1.67 $\times 10^{-1}$  &  2.24 $\times 10^{-3}$ &  2.70 $\times 10^{-3}$ \\
0.9  &  8.53 $\times 10^{-1}$ &  8.53 $\times 10^{-1}$  &  7.06 $\times 10^{-3}$ &  8.16 $\times 10^{-3}$ \\
1.0  &  3.63 $\times 10^{0}$ &  3.63 $\times 10^{0}$  &  1.75 $\times 10^{-2}$ &  1.98 $\times 10^{-2}$ \\
1.25  &  6.65 $\times 10^{1}$ &  6.65 $\times 10^{1}$   &  9.22 $\times 10^{-2}$ &  9.97 $\times 10^{-2}$ \\
1.5  &  5.41 $\times 10^{2}$ &  5.41 $\times 10^{2}$  &  2.89 $\times 10^{-1}$ &  3.05 $\times 10^{-1}$ \\
1.75  &  2.52 $\times 10^{3}$ &  2.52 $\times 10^{3}$   &  6.77 $\times 10^{-1}$ &  7.06 $\times 10^{-1}$ \\
2.0  &  8.10 $\times 10^{3}$ &  8.10 $\times 10^{3}$  &  1.30 $\times 10^{0}$ &  1.30 $\times 10^{0}$ \\
2.5  &  4.10 $\times 10^{4}$ &  5.01 $\times 10^{4}$  &  3.34 $\times 10^{0}$ &  4.16 $\times 10^{0}$ \\
3.0  &  1.18 $\times 10^{5}$ &  1.88 $\times 10^{5}$  &  6.19 $\times 10^{0}$ &  9.58 $\times 10^{0}$ \\
3.5  &  2.47 $\times 10^{5}$ &  5.17 $\times 10^{5}$  &  9.47 $\times 10^{0}$ &  1.82 $\times 10^{1}$ \\
4.0  &  4.22 $\times 10^{5}$ &  1.14 $\times 10^{6}$  &  1.27 $\times 10^{1}$ &  3.05 $\times 10^{1}$ \\
5.0  &  8.54 $\times 10^{5}$ &  3.71 $\times 10^{6}$  &  1.86 $\times 10^{1}$ &  6.77 $\times 10^{1}$ \\
6.0  &  1.30 $\times 10^{6}$ &  8.61 $\times 10^{6}$  &  2.28 $\times 10^{1}$ &  1.22 $\times 10^{2}$ \\
7.0  &  1.72 $\times 10^{6}$ &  1.63 $\times 10^{7}$  &  2.56 $\times 10^{1}$ &  1.93 $\times 10^{2}$ \\
8.0  &  2.05 $\times 10^{6}$ &  2.71 $\times 10^{7}$  &  2.72 $\times 10^{1}$ &  2.79 $\times 10^{2}$ \\
9.0  &  2.32 $\times 10^{6}$ &  4.06 $\times 10^{7}$  &  2.79 $\times 10^{1}$ &  3.79 $\times 10^{2}$ \\
10.0  &  2.51 $\times 10^{6}$ &  5.76 $\times 10^{7}$   &  2.81 $\times 10^{1}$ &  4.90 $\times 10^{2}$
\end{longtable}

\bibliographystyle{apsrev}

\end{document}